\documentclass{aa}
\usepackage[utf8]{inputenc}
\usepackage{graphicx}
\usepackage{longtable}
\usepackage{rotating}
\usepackage[normalem]{ulem}
\usepackage{amsmath}
\usepackage{amssymb}
\usepackage{caption}
\usepackage{subcaption}
\usepackage{siunitx}
\usepackage{hyperref}
\usepackage[varg]{txfonts}
\usepackage{natbib}
\bibpunct{(}{)}{;}{a}{}{,} 
\usepackage{booktabs}
\hypersetup{
    colorlinks=true,
    citecolor=blue,
    urlcolor=blue,
}
\author{
    T.~E.~Rivera-Thorsen \thanks{\email{\url{trive@astro.su.se}}}
    \and
    M. Hayes
    \and
    J. Melinder
}
\authorrunning{T.~E.~Rivera-Thorsen et al.}
\titlerunning{Lyman Continuum escape in the UVUDF}
\institute{The Oskar Klein Centre, Department of Astronomy,
          Stockholm University, AlbaNova, SE-10691 Stockholm, Sweden
}
\abstract
{When studying the production and escape of Lyman Continuum from galaxies, it is 
  standard to rely on array of indirect observational tracers in preselection of 
  candidate leakers.}
{In this work, we investigate how much ionizing radiation might be missed due to 
  these selection criteria by completely removing them and performing a search 
  selected purely from rest-frame LyC emission; and how that affects our 
  estimates of the ionizing background.}
{We invert the conventional method and perform a bottom-up search for 
  Lyman-continuum leaking galaxies at redshifts $2\lesssim z \lesssim 3.5$. 
  Using archival data from HST and VLT/MUSE, we run source finding software on 
  UV-filter HST images from the HUDF, and subject all detected sources to a 
  series of tests to eliminate those that are inconsistent with being ionizing
  sources.}
{We find 6 new and one previously identified candidate leakers with absolute
  escape fractions ranging from 36\% to $\sim 100\%$. Our filtering criteria 
  eliminate one object previously reported as a candidate ionizing emitter in 
  the literature, while we report non-detection in the rest frame Lyman 
  continuum of two other previously reported sources. We find that our 
  candidates make a contribution to the metagalactic ionizing field of 
  $\log_{10}(\epsilon_{\nu}) = 25.32^{+0.25}_{-0.21}$ and 
  $25.29^{+0.27}_{-0.22}$ erg $s^{-1}$ Hz$^{-1}$ cMpc$^{-3}$ for the full set of 
  candidates and for the 4 strongest candidates only; both values are higher than 
          but consistent with other recent figures in the literature.}
{Our findings suggest that galaxies that do not meet the usual selection 
  criteria may make a non-negligible contribution to the cosmic ionizing field. 
  We recommend that similar searches be carried out on a larger scale in 
  well-studied fields with both UV and large ancillary data coverage, for 
  example in the full set of CANDELS fields.}


\usepackage{booktabs}

\date{}
\title{A bottom-up search for Lyman-continuum leakage in the Hubble Ultra Deep Field}
\hypersetup{
 pdfauthor={Emil Rivera-Thorsen},
 pdftitle={A bottom-up search for Lyman-Continuum leakage in the Hubble Ultra Deep Field},
 pdfkeywords={},
 pdfsubject={},
 pdfcreator={Emacs 27.2 (Org mode 9.5.2)},
 pdflang={English}}
\begin{document}

\maketitle

\section{Introduction} \label{sec:orge4ffce6} The Epoch of Reionization (EoR)
was the last major phase transition of the Universe, and ended around z \(\sim\)
5.5 - 6 \citep[e.g.,][]{fan2006,dayal2020,bosman2022}. At the relevant
redshifts, the Cosmic AGN population was probably still too low to dominate the
metagalactic ionizing radiation
\citep[e.g.,][]{haardt2012,fauchergiguere2020,madau2015}. Young, hot stars in
star forming galaxies thus most likely contributed the bulk of ionizing photons.
Balancing the budget of ionizing photons required to account for reionization is
still an ongoing topic of research. The amount of ionizing photons required to
account for reionization is still debated \citep[e.g.,][]{davies2021}, as is the
time scale and thus the required photon production rate \citep[e.g.,][and
references therein]{naidu2020a,becker2021}. In order to account for the EoR, an
estimated fraction of the produced ionized photons \(f_{\text{esc}}^{LyC}
\approx\) 10 -- 20 \% must escape their source galaxies
\citet{robertson2015,finkelstein2019,naidu2020a}, yet in the low- and
intermediate redshift Universe, neither the number of LyC-emitting galaxies nor
their average escape fractions meet this requirement. In the local Universe,
Lyman Continuum Emitters (LCEs) seem to be exceedingly rare, with just over 50
confirmed leakers known at \(z < 0.5\)
\citep{bergvall2006,leitet2011,leitet2013,borthakur2014,leitherer2016,izotov2016,
izotov2016nature,izotov2018a,izotov2018b,wang2019,malkan2021b,flury2022a}. 
Earlier studies find escape fractions of only a few percent, but later searches
focusing on extreme emission line galaxies have yielded a much wider range of
line-of-sight escape fractions
\citep{izotov2016,izotov2016nature,izotov2018a,izotov2018b}. At ``cosmic noon'',
2 \(\le\) \emph{z} \(\le\) 4, a roughly similar number of confirmed leakers have
been found \citep{vanzella2012,mostardi2015,shapley2016,debarros2016,
        vanzella2016,bian2017,fletcher2018,chisholm2018a,steidel2018,vanzella2018,
riverathorsen2019,ji2020,marqueschaves2021}, as well as a number of candidates
still awaiting follow-up and confirmation. Only one confirmed leaker has been
found in the difficult window of \(1 \lesssim z \lesssim 2\), using the
\emph{Astrosat} space observatory \citep{saha2020}. In addition, stacking
analyses at redshifts 1 \(\le\) \emph{z} \(\le\) 3 have shown the overall cosmic
escape fraction to be low at redshifts < 3. \citet{cowie2009} found a
2-\(\sigma\) upper limit to the average escape fraction of < 0.8\% at 0.9 <
\emph{z} < 1.4 using stacked \emph{GALEX} imaging, concluding that galaxies can
only have accounted for reionization if the average escape fraction evolves
strongly between redshifts 1.4 and 5. For a similar redshift range,
\citet{alavi2020} found a 3-\(\sigma\) limit of the average escape fraction at
\(z \sim 1.3\) of < 7\%, although this was for a relatively small sample,
selected specifically to be young, highly star forming, low-mass and
low-metallicity; and \citet{rutkowski2016} found an upper limit to the escape
fraction of star forming galaxies at \(z \sim 1\) to be f\textsubscript{esc} <
2.1\%, and of f\textsubscript{esc} < 9.6\% for a subsample selected to have
W\textsubscript{H\(\alpha\)} 200 Å, which are believed to be the closest
intermediate redshift analogs to star-forming galaxies at redshifts \(z \ge 6\).

At higher redshifts, \citet{grazian2016,grazian2017} find from a stacking
analysis in the GOODS-N, EGS and COSMOS fields an average f\textsubscript{esc}
\(\lesssim 2\%\) for bright galaxies and \(\lesssim 10\%\) for faint galaxies at
\(z=3.3\). They conclude that the ionizing output in their search volume is too
low by a factor of 2 to account for the ionized state of the Universe at that
redshift and insufficient to account for the epoch of reionization without a
significant evolution in average escape fraction.

The exact mass, luminosity and number distribution of Lyman Continuum (LyC)
leaking galaxies cannot be studied directly at redshifts \(\gtrsim 4\), at which
point the neutral fraction of the intergalactic medium (IGM) becomes high enough
to effectively make it opaque to ionizing radiation \citep{inoue2014}. To study
the processes which govern LyC escape, and in turn the process of Reionization,
we need reliable tracers of escaping LyC, and reliable estimators of intrinsic
LyC. The production rate is readily determined from observable quantities; but
the escape depends on a complex interplay of ISM conditions including star, gas
and dust content, ISM geometry, kinematics and ionization parameter and
metallicity.

A popular candidate tracer is the ratio of doubly to singly ionized Oxygen, as
measured in the forbidden optical lines, [\ion{O}{iii}]/[\ion{O}{ii}] or
O\textsubscript{32} \citep{jaskot2013,nakajima2014,keenan2017}. The strength of
the [\ion{O}{iii}] \(\lambda \lambda\) 4960,5008 Å doublet reflects the ionizing
output of the stars in the region, and the relative strengths of the
[\ion{O}{iii}] and [\ion{O}{ii}] traces the relative amounts of doubly and
singly ionized oxygen and thus indirectly the ionization properties of the ISM
surrounding the LyC producing stellar region. The ratio is easily observed but
as a predictor of f\textsubscript{esc,LyC} it has shown mixed results
\citep{jaskot2019,bassett2019,nakajima2020}. Still, it remains a widely used
tool for candidate preselection.

In addition to serving as tracers of ionizing emissivity during the EoR and
subsequent epoch when LyC cannot be observed directly, these observables also
provide efficient and convenient preselection criteria when carrying out
searches for LyC leakers in a given cosmic volume; and indeed every search
listed above rely on either in the literature at the time of writing this has
relied of one or more of these preselection criteria. Each of these criteria
however has some associated selection bias. The equivalent width of H\(\alpha\),
which traces strong specific star formation, was a applied as the main selection
criterion by e.g. \citet{rutkowski2016}. O\textsubscript{32} is a tracer of
compound effects of star formation and ISM ionization and has been widely
applied as a preselection criterion, among others by
\citet{rutkowski2017,flury2022a,izotov2016,izotov2016nature,izotov2018a,izotov2018b}.
Photometric rest-frame UV colors \citep{flury2022a}, including Lyman Break
selection \citep{steidel2018}, is cheap and effective but only traces observed
UV photon production, and LBG selection criteria directly exclude galaxies with
large f\textsubscript{esc,LyC} \citep{cooke2014}.

\citet{rutkowski2017} performed a stacking analysis of emission line selected
galaxies at \(z \sim 2.5\), and found that the escape fraction for [\ion{O}{ii}]
selected galaxies was \(\lesssim 5.6\%\), and \(\lesssim 14.0\%\) for galaxies
selected for [\ion{O}{iii}]/[\ion{O}{ii}] \(\ge\) 5, not enough to cause and
maintain ionization of the IGM unless the number of these extreme emission line
galaxies grows substantially between redshifts 2 and 6.

Lyman-\(\alpha\) line properties is another widely used preselection criterion
\citep[e.g.][]{fletcher2018,riverathorsen2019}. Ly\(\alpha\) radiative transfer
is regulated by and highly sensitive to the same \ion{H}{i} which also absorbs
LyC, although the effects are different and the comparison thus not
straightforward \citep{verhamme2015,behrens2014,kakiichi2021}. Strong
Ly\(\alpha\) emission with a narrow, double-peaked line profile with narrow peak
separation has proven a reliable, of not necessarily complete, predictor of LyC
emission \citep{izotov2018a,kakiichi2021,flury2022a}. It however is unclear how
much of Cosmic LyC is traced by Ly\(\alpha\). The majority of ionizing photons
in the Universe at \(z \sim 3\) are produced in Lyman Break galaxies or local
analogs, but only \(\sim\) 20\% of LBGs emit strongly in Ly\(\alpha\)
\citep{shapley2003,stark2010}. The average Ly\(\alpha\) escape fraction of \(z
\sim 3\) LBGs is 5-10\%, but quickly increasing towards higher redshifts
\citep{hayes2011}. It is possible that LyC escape may occur from galaxies with
faint or undetectable Ly\(\alpha\) emission.

In this work, we mitigate the potential biases introduced by such preselection
by carrying out a search for any UV source we could find in archival data of the
Hubble Ultra Deep Field \citep[HUDF,][]{beckwith2006}; and for each detected UV
source using ancillary data to determine whether it is likely to be a
LyC-leaking galaxy within the redshift range appropriate for the detection
filter. In essence, one could say that this is a \emph{Lyman-continuum selected}
survey and should capture every LyC-leaking source in the field, regardless of
other properties. Of course, the overwhelming majority of sources detected in
these filters is expected to be non-ionizing radiation from lower-redshift
galaxies, but current datasets have reached a point where the completeness of
spectroscopic catalogs delivered by large-format IFUs is sufficient to provide
redshifts for the great majority of these sources.

We perform the source detection in archival \emph{Hubble} WFC3/UVIS UV imaging
of the HUDF \citep[The \emph{UVUDF},][hereafter
R15]{teplitz2013,rafelski15_uvudf}, and for each source, extract fluxes in a
small image segment at its location on the sky in the other UVUDF filters, as
well as the original optical HUDF ACS filters, and the WFC3/IR filters of
\citet{oesch2010b,bouwens2011,ellis2013}. Furthermore, we rely on the extensive
spectroscopic redshift catalog for this field provided by \citet[][hereafter
I17]{inami2017} as part of the larger MUSE-Deep survey \citep{bacon2015} to
determine the redshift of our sources, where a catalog entry exists.

Recently, two projects have surveyed larger fields containing the HUDF in
redshift ranges overlapping with this work \citep{jones2021,saxena2021}; both
relying on preselection steps to identify candidates for deeper investigation.
This provides an excellent opportunity for direct comparison with these methods
and a deeper discussion of their biases, advantages and drawbacks.

The rest of this paper is structured as follows. In Sect.~\ref{sec:data}, we
briefly describe the data on which this work builds. In
Sect.~\ref{sec:analysis}, we describe our methodology, the various steps in our
process of selecting candidate objects, and present the final selection. In
Sect.~\ref{sec:candchar}, we present a physical characterization of the
candidate galaxies, and we discuss the methods and results in
Sect.~\ref{sec:discuss}. Finally, we give a summary of our findings in
Sect.~\ref{sec:summary}. Throughout this paper, we assume a standard concordance
flat \(\Lambda\)-CDM cosmology with H\textsubscript{0}=70 km/s/Mpc,
\(\Omega\)\textsubscript{\(\Lambda\)}=0.7, and \(\Omega\)\textsubscript{m}=0.3.
Magnitudes are given in the AB system.

\section{Data                                             \label{sec:data}}
\label{sec:orgb7066e8}
The purpose of this study is to test how much escaping ionizing radiation, if
any, is not accounted for by the most common preselection methods. Instead, we
have performed an unbiased source detection based on UV imaging data from the
HST, selecting interesting and candidate objects through a series of tests and
visual inspections as described below.

Imaging data used in this paper are the same as those presented in
\citet{teplitz2013} and \citet{rafelski15_uvudf} (R15 in the following), and
extensive information about observations and reductions can be found in these
two papers. Here we present only a brief summary. We adopt the full wavelength
range of HST imaging for the UDF: The F225W, F275W, and F336W data obtained from
WFC3/UVIS under the UVUDF program, the F435W, F606W, F775W, and F850LP imaging
obtained with ACS/WFC under the original HUDF campaign \citep{beckwith2006}, and
the F105W, F125W, F140W, and F160W frames obtained with WFC3/IR under both the
UDF09 and UDF12 campaigns
\citep{oesch2010b,bouwens2011,ellis2013,koekemoer2013}. We note that the UVIS
imaging obtained here falls within the footprint of the ACS optical imaging, but
only \(\approx\) 80~\% of the survey area is covered by the IR imaging. IR data
can greatly help determine the redshifts and physical properties of the stellar
population, but where none was present, we have still included objects with UV
and optical colors suggesting them to be possible LyC leakers.

Using an array of different filters, the relative photometry will be affected by
the relative point-spread function at each wavelength of each camera. All PSFs
were created by stacking stars, while for both WFC3 UVIS and IR data the PSFs
were supplemented by models PSFs that describe the outer wings; see R15 for more
details. Matching and convolution were then performed using the \texttt{psfmatch} and
\texttt{fconvolve} tasks in \texttt{IRAF}.
In addition, we have extracted spectra from MUSE datacubes from the official
data products released by the MUSE-DEEP \citep{bacon2015}, made available from the
ESO archives.\footnote{\url{https://archive.eso.org/cms/eso-archive-news/release-of-pipeline-processed-muse-deep-3d-data-cubes.html}}

\begin{figure}[htb]
\centering
\includegraphics[width=\columnwidth]{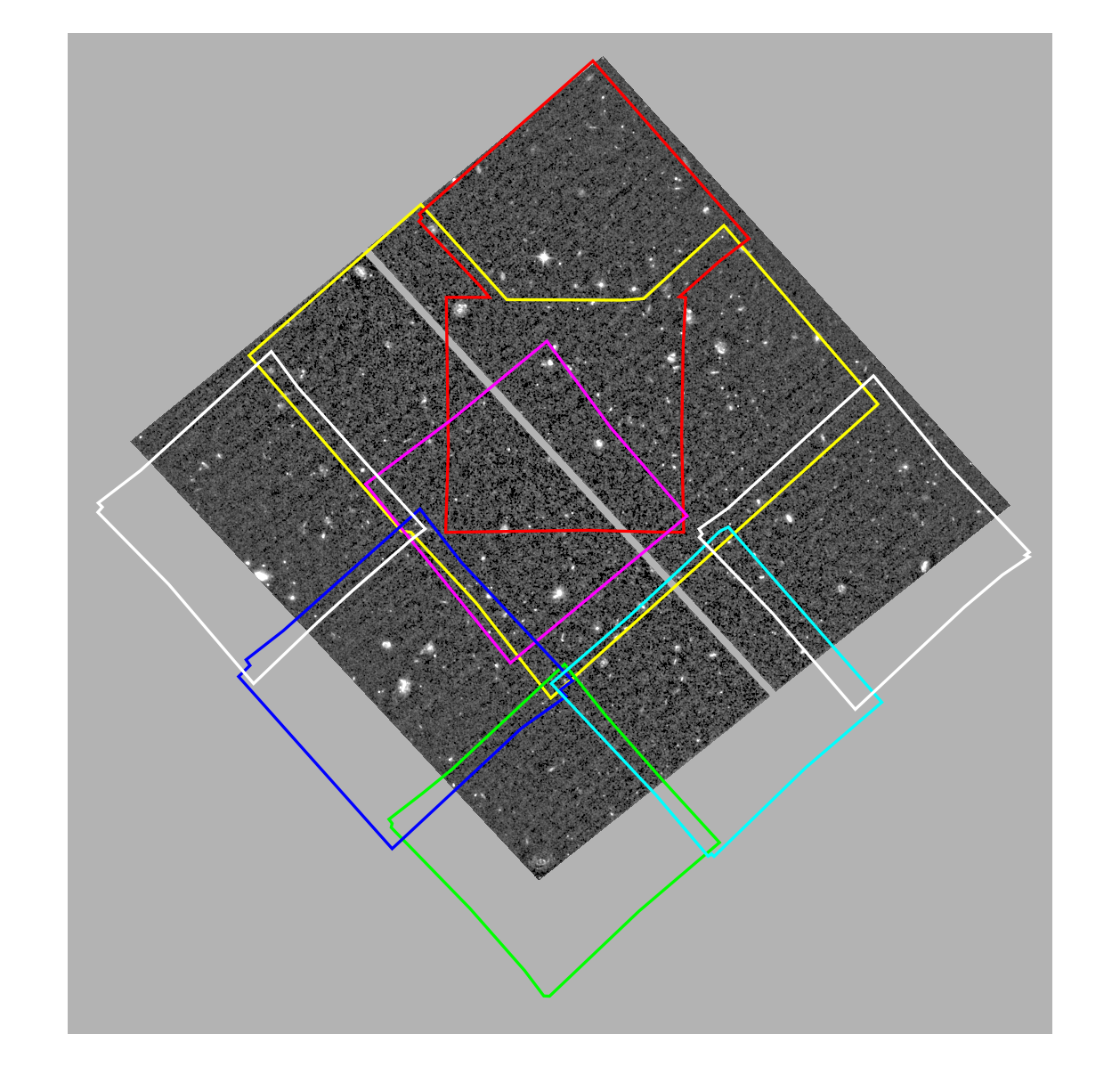}
\caption{\label{fig:MUSEfootprints}Approximate footprints of the combined MUSE-Deep cubes superimposed on the UVUDF FOV, represented by the PSF matched F275W frame. Colors to help distinguish individual MUSE pointings and mosaics.}
\end{figure}

Fig.~\ref{fig:MUSEfootprints} shows the footprints of the publicly available
MUSE-DEEP datacubes overlaid on the PSF-matched and background subtracted UVUDF
F225W image. The overlap is large but not complete, and we did find a few
candidates in the UVUDF falling outside the MUSE-DEEP footprint.

\section{Analysis and results                             \label{sec:analysis}}
\label{sec:org9d3ee29}

\subsection{Strategy overview} \label{sec:org7cd6baa}

Here, we give a quick overview of the search strategy; each step will be
described in greater depth below. In a nutshell, our strategy can be formulated
as: \emph{Detect every source present in a given UV filter, and subsequently use
images in additional, redder, filters as well as ancillary information to
determine for each source whether it is consistent with being escaping LyC
radiation in the detection filter.} To this end, we first ran source finding and
detection routines on each of the three background subtracted
(Sect.~\ref{sec:orgf584df0}) UV filter images using a set of settings which
ensure that we catch as many faint but real sources at the cost of yielding a
potentially substantial number of spurious detections
(Sect.~\ref{sec:org468fc43}).

\begin{figure}[tbh]
  \centering
  \includegraphics[width=\columnwidth]{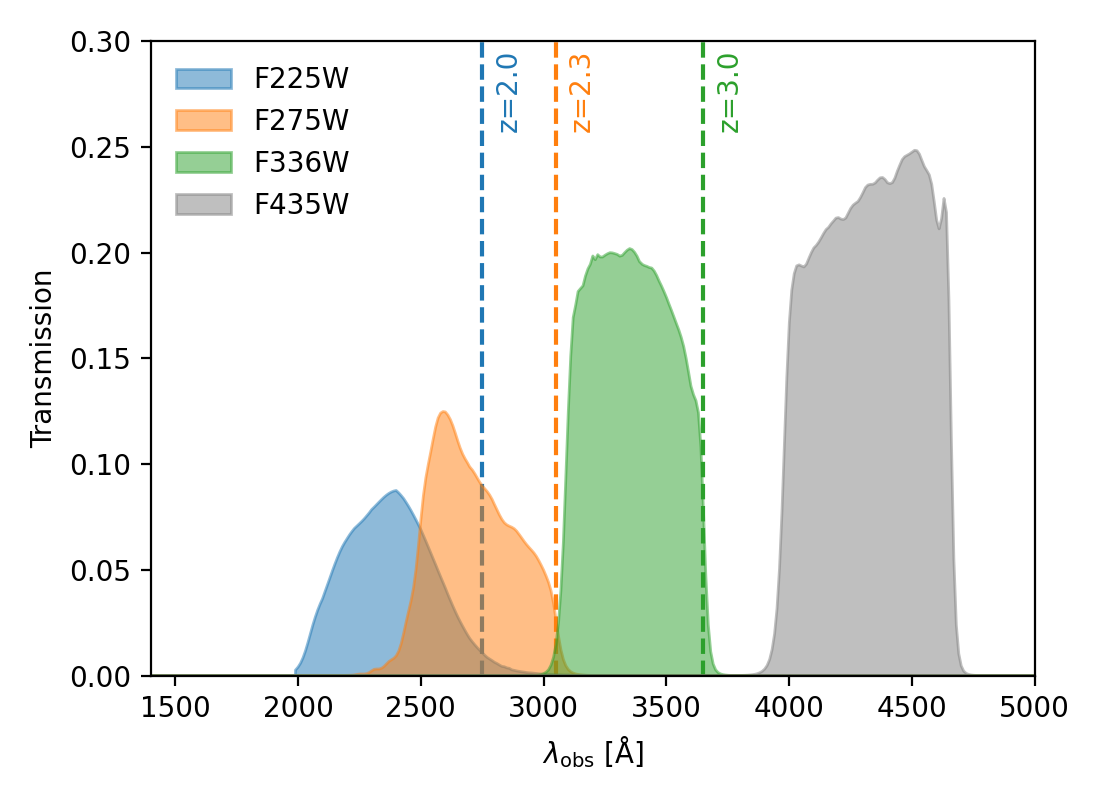}
  \caption{\label{fig:filtersandzs}Filter throughput curves for the three
    detection filter plus F435W. For each detection filter, we also show the
    approximate lower redshift cut-off for galaxies to be
    reliably detected in LyC. Verification of detections in any filter is
    done in the filter immediately to the right.
    }
\end{figure}

Using the segmentation maps created in the source detection step, we extracted
fluxes for each segment in each of the eleven filters (or the available subset
hereof, Sect.~\ref{sec:mcerrs}). For each filter, the adjacent filter to the red
side was used as a verification filter; any detection that was not also detected
in the verification filter was immediately rejected
(Sect.~\ref{sec:org90016a9}). Since we are searching for objects at a redshift
which puts LyC in the detection filter and nonionizing UV continuum in the
verification filter, we also required that an object must be of lower
\(m_{\text{AB}}\) in the verification filter than the detection filter
(Sect.~\ref{sec:uvcolors}). Figure~\ref{fig:filtersandzs} shows throughput
curves of the three UV filters and, for each of them, its pivot wavelength and
the redshift at which the Lyman edge of 912 \AA{} falls approximately at its red
edge. Also shown is the optical filter F435W, used for verification of sources
detected in F336W (see below).

We then moved on to visual inspection of the remaining segments for morphology
and for their SED, to remove objects which are clearly not LyC leaking galaxies
at redshift 1.9--3.5 (Sect.~\ref{sec:morphvis}). Next, we matched the remaining
objects to the catalogs of R15 and I17 for existing photometric or spectroscopic
redshifts. We also performed our own SED fitting using \texttt{BAGPIPES}
\citep{carnall2018}, and extracted spectra from the MUSE-DEEP public datacubes
for selected objects. Where present and of medium or high certainty (see
Sect.~\ref{sec:org0d21d3b}), spectroscopic redshifts were given priority. Where
only photometric redshifts were present, we made a case-by-case assessment. Our
final list of 7 candidate objects is tabulated in Table~\ref{tab:baseprops},
which lists their sky coordinates as well as the catalog IDs of the
corresponding objects in the catalogs of R15 and I17.

We finally re-measured fluxes for the remaining candidates in all available
filters integrated in a circular aperture with a fixed diameter of \(0\farcs 8\)
(Sect.~\ref{sec:faphot}). This aperture size is equal to the value of the seeing
adopted in I17, and we found that it encompasses the majority of flux from our
remaining candidates, unlike the often small image segments found from source
detection in the rest-frame FUV. These fluxes enabled us to find integrated
stellar population properties which in turn allows for a more direct comparison
to other galaxies in the literature.

\begin{table}[tbh]
\begin{center}
  \caption{\label{tab:baseprops}Candidate objects coordinates and basic
  properties.}
\begin{tabular}{{lrrrr}}
\toprule
ID & RA & DEC & ID\textsubscript{I17} & ID\textsubscript{R15}\\
 & J2000 & J2000 &  & \\
\midrule
F275W--314 & 53.168794 & -27.796954 & 1270 & 3686\\
F275W--2055 & 53.154158 & -27.759680 & -- & 6958\\
\textbf{F336W--189} & 53.167905 & -27.797922 & 1087 & 3506\\
\textbf{F336W--554} & 53.136844 & -27.786032 & 1249 & 6549\\
F336W--606 & 53.153802 & -27.784799 & -- & 37178\\
\textbf{F336W--1013} & 53.173526 & -27.773692 & 7380 & 24800\\
F336W--1041 & 53.169621 & -27.772601 & 138 & --\\
\textbf{F336W--1041} & 53.169621 & -27.772601 & 6292 & 24755\\
\bottomrule
\end{tabular}
\tablefoot{Center coordinates of the UV detections
    of candidate objects, as well as catalog IDs of counterparts in the R15 and
    I17 catalogs. Bold text denotes our tier 1 candidates (see below).
    F336W--1041 has two matches in the I17 catalog within the \(0\farcs 75\)
    tolerance; we have not listed the R15 counterpart of the I17 matched object
    with the largest angular separation from the source.}
\end{center}
\end{table}

\begin{table*}[tbhp]
  \caption{\label{tab_redshifts} Photometric and spectroscopic
  candidate redshifts.}
\centering
\begin{tabular}{{lrrrrr}}
\toprule
ID & z\textsubscript{MUSE} & z\textsubscript{I17}\tablefootmark{a} & z\textsubscript{Bagpipes} & BPZ\tablefootmark{\(b\)} & z\textsubscript{EAZY}\tablefootmark{\(b\)}\\
 & (3) & (4) & (5) & (6) & (7)\\
\midrule
F275W--314 & 1.995 & 1.995011 (3) & 2.09 $\pm$ 0.08 & 2.073 (0.99) & 1.825 (1.00)\\
F275W--2055 & -- & -- & 2.07 $\pm$ 0.08 & 2.430 (0.99) & 2.256 (1.00)\\
\textbf{F336W--189} & 3.46 & 3.462787 (3) & 0.11 $\pm$ 0.05 & 3.313 (1.00) & 0.173 (1.00)\\
\textbf{F336W--554} & -- & 2.867758 (3) & 0.47 $\pm$ 0.05 & 2.972 (1.00) & 2.960 (1.00)\\
F336W--606 & -- & -- & 2.41 $\pm$ 0.15 & 2.69 (0.99) & 2.50 (1.00)\\
\textbf{F336W--1013} & 2.98 & 2.981311 (2) & 2.04 $\pm$ 0.07 & 1.893 (0.99) & 1.811 (1.00)\\
F336W--1041 & -- & 3.330044 (2) & 2.19 $\pm$ 0.09 & -- & --\\
\textbf{F336W--1041} & 3.329 & 3.329164 (2) & 2.19 $\pm$ 0.09 & 2.186 (0.99) & 2.004 (1.00)\\
\bottomrule
\end{tabular}
\tablefoot{Redshifts calculated by various methods for objects of interest in our sample. \texttt{BAGPIPES} -computed 
    redshifts are based on segment fluxes; updated \texttt{BAGPIPES} redshifts 
    for F275W-2055 and F336W-606 based on larger apertures are reported below.\\
    \tablefoottext{a}{ Numbers i parentheses are the confidence levels from
        I17 on a scale from 1-3, with higher being better.}\\
    \tablefoottext{b}{Numbers in parentheses are the integrated probabilities 
        (\texttt{ODDS}) reported by R15, where they are also explained (p. 9). 
        According to \citeauthor{rafelski15_uvudf}, a value of \(\ge\) 0.9 is 
        considered a high-probability determination.}}
\end{table*}

\subsection{Background subtraction}
\label{sec:orgf584df0}
We modeled and subtracted the backgrounds of all drizzled and PSF matched images
using the python \texttt{photutils} package \citep{photutils2020}, using the
\texttt{Background2D} class, which estimates the background locally on a mesh,
based on sigma clipping statistics, and subsequently interpolates this
background mesh to the resolution of the original image. We used the default
\texttt{SExtractorBackground} background estimator and \texttt{StdBackgroundRMS}
background RMS estimator.

Following the recommendations from the \texttt{photutils} authors, we estimated
the background in two steps. In the first step, the background was estimated and
subtracted on the unmasked images. We then ran the \texttt{make\_source\_mask}
function, which uses sigma clipping statistics for an initial noise and
background estimate, then uses this to detect sources, create a segmentation
map, dilate the source segments, and finally output a mask array. This mask was
then passed to a second iteration with the \texttt{Background2D} class, along
with the unmodified data frames, and the resulting background image subtracted
from the latter.

The UV frames contain quite complex and strong features, what the authors of R15
describe as a ``blotchy pattern'' which, given the relatively weak signals and
strongly correlated noise patterns, could lead to over- or underestimation of
flux in certain regions, in turn bringing detected flux below the
signal-to-noise threshold or even below zero, if not properly modeled and
subtracted. It was therefore necessary to choose a \texttt{box\_size} small
enough to encompass the characteristic size of these blotches, in order to model
these residual patterns, yet large enough to not include noise in the model.

In the optical/red and IR filters, extended red stellar populations become
visible in the field's elliptical galaxies. These pose a challenge for
background subtraction, as they require either large box sizes or strong mask
dilation in order to avoid subtracting the outer stellar halos. We opted for a
combination of the two, although with the main emphasis on larger box size,
since the background in the optical and red filters has less structure than in
the UV, and since there are so many objects visible in the red and IR filters
that dilating the masks would unnecessarily remove a large fraction of the
background pixels, severely impacting the quality of background modeling. We
note that due to the fact that we target relatively high surface brightness and
compact sources, this will make only a small difference to the final photometry
in the redder bands.

In Appendix~\ref{app.bgsub}, we show examples of background and background
subtracted data in F225W and F775W to illustrate the different challenges
presented by the UV and red data.
Parameters passed to the source mask creation and background modeling routines
for each filter were determined by trial and error, visually inspecting the
modeled background for over- or undersubtracted regions near masked sources, and
artifacts of poorly modeled background. The parameters that we finally settled
on are tabulated in Table \ref{tab:bgsubtpars}.

\begin{figure}[tbh]
\begin{center}
\includegraphics[width=\columnwidth]{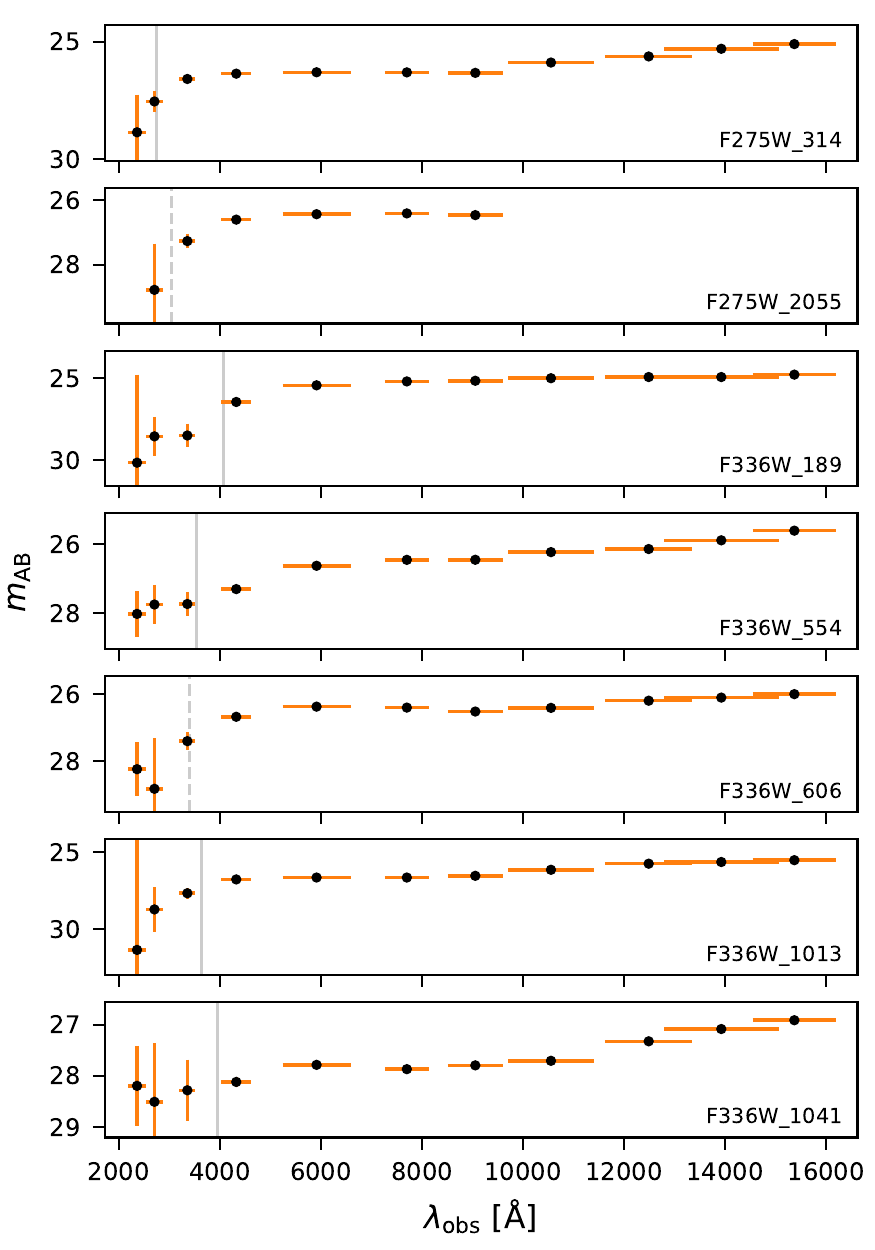}
\caption{\label{fig:leakerSEDs}Spectral Energy Distributions of the leaker
    candidates, extracted in circular apertures of radius \(0 \farcs 4\). The
    dotted vertical lines mark the Lyman Edge at the redshifts listed in
    Table~\ref{tab:bpout_fullapert_fixmet}. Note that errors in the UV bands are
    inflated for the reasons described in  \ref{sec:mcerrs} and \ref{sec:faphot}.
    Some datapoints here are shown as such despite being consistent with zero;
    see Fig.~\ref{fig:sedfig} for a comparison of uncertainties in segments and 
    fixed apertures.}
\end{center}
\end{figure}

\subsection{Source detection and deblending}
\label{sec:org468fc43}

We first ran the \texttt{photutils} source detection routines
\texttt{detect\_sources()} with settings which prioritized completeness in
detection over avoiding spurious detections, since these as discussed below will
be rejected in later steps. We then passed the resulting segmentation maps to
the function \texttt{deblend\_sources()} to deblend overlapping galaxies. In the
latter step, we again allowed the algorithms to deblend too many rather than too
few sources, since LyC escape is a local phenomenon which will not suffer any
missed detections due to this, while larger apertures, as discussed in
Sect.~\ref{sec:faphot}, may dilute the SNR of a detected source below our
threshold value. This resulted in a deblended segmentation map for each of the
three UV filters. For each of these, we then extracted fluxes in all segments in
every one of the remaining filters, including the two remaining UV filters.
Based on SNR and color selection criteria described below, we could then reject
objects that are inconsistent with being Lyman Continuum leakers detected in the
ionizing continuum.

Working on PSF-matched data comes at a cost in terms of image depth, since the
UV and optical data are convolved and mixed with a number of pixels which
contribute only noise, degrading the signal-to-noise ratios of fluxes measured
in these images. \citeauthor{rafelski15_uvudf} have opted to optimize their
photometric depth by measuring all fluxes in unconvolved data, and subsequently
applying an aperture loss correction in the NIR data, where the PSF is
significantly larger than in the UV/optical bands. However, such an approach
would result in practical difficulties in this work, rendering it unfeasible.
Most importantly, such an aperture correction accounts for flux lost outside
aperture due to convolution with the PSF, but not for flux from outside sources
bleeding into the aperture. When using the NUV bands for detection images, a
given segment used for photometry may only cover a small part of a galaxy and in
general not the part that is brightest in the optical and NIR bands, and thus
flux from neighboring regions may bleed into the segment in bands with larger
PSF. Thus, when performing photometry on segments of galaxies, this must be done
on PSF matched images.

\begin{table*}[htbp]
  \caption{\label{tab:fluxes}Segment-measured AB magnitudes for the candidate
    objects in the five bluest filters in the UVUDF.}
\centering
\begin{tabular}{{lccccc}}
\toprule
Filter & F225W & F275W & F336W & F435W & F606W\\
Object &  &  &  &  & \\
\midrule
F275W-314 & 30.0 $\pm$ 0.7 & 29.12 $\pm$ 0.25 & 28.10 $\pm$ 0.10 & 27.822 $\pm$ 0.028 & 27.767 $\pm$ 0.018\\
F275W-2055 & --- & 30.6 $\pm$ 0.4 & 29.58 $\pm$ 0.10 & 28.97 $\pm$ 0.05 & 28.785 $\pm$ 0.027\\
\textbf{F336W-189} & 29.8 $\pm$ 0.5 & 29.3 $\pm$ 0.5 & 29.47 $\pm$ 0.30 & 28.160 $\pm$ 0.033 & 27.250 $\pm$ 0.010\\
\textbf{F336W-554} & 29.80 $\pm$ 0.27 & 30.1 $\pm$ 0.5 & 29.73 $\pm$ 0.26 & 29.60 $\pm$ 0.10 & 29.05 $\pm$ 0.04\\
F336W-606 & 29.3 $\pm$ 0.5 & 29.4 $\pm$ 0.5 & 28.82 $\pm$ 0.24 & 28.40 $\pm$ 0.05 & 28.049 $\pm$ 0.024\\
\textbf{F336W-1013} & --- & 29.7 $\pm$ 0.9 & 28.13 $\pm$ 0.15 & 27.560 $\pm$ 0.021 & 27.470 $\pm$ 0.014\\
\textbf{F336W-1041} & 30.3 $\pm$ 0.4 & 30.9 $\pm$ 0.8 & 30.04 $\pm$ 0.32 & 29.86 $\pm$ 0.11 & 29.71 $\pm$ 0.07\\
F336W-1043 & 30.2 $\pm$ 0.5 & 29.59 $\pm$ 0.31 & 29.93 $\pm$ 0.31 & 32.7 $\pm$ 1.6 & 31.5 $\pm$ 0.4\\
\bottomrule
\end{tabular}
\end{table*}

\begin{figure}[htb]
\begin{center}
\includegraphics[width=\columnwidth]{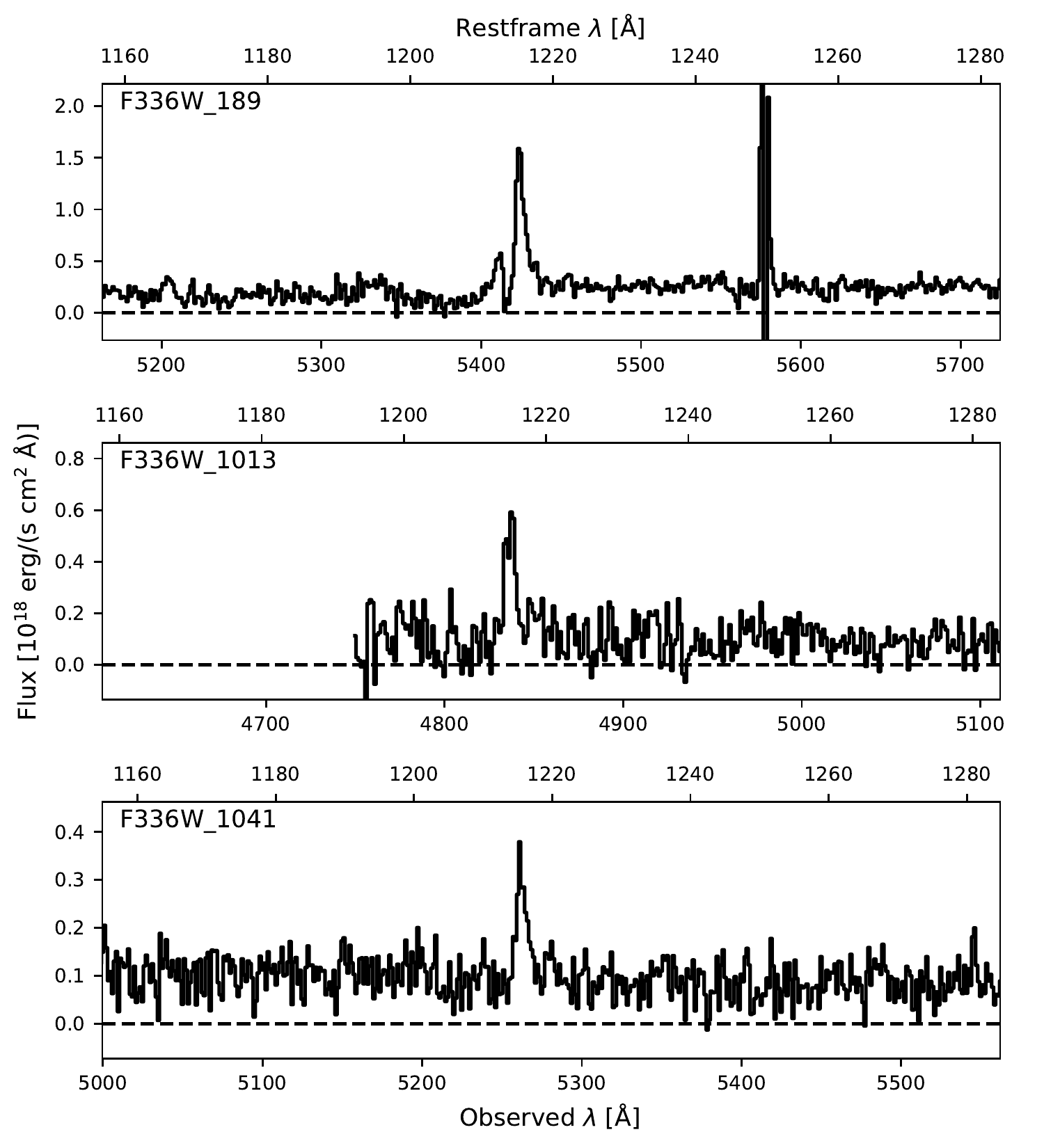}
\caption{\label{fig:OOIspecs}Spectral regions around rest-frame Ly\(\alpha\) of
  the subset of candidate objects with Ly\(\alpha\) in emission. At the bottom
  is shown the observed wavelength scale, above each panel is the corresponding
  rest-frame wavelength range.}
\end{center}
\end{figure}

\subsection{Photometry and uncertainties                    \label{sec:mcerrs}}
\label{sec:orge0b4a7e}
For each or the three UVUDF filters, we created a segmentation image as
described above, and proceeded to measure fluxes in all segments of this image
for every one of the filters listed in Table~\ref{tab:bgsubtpars}, resulting
in a measured flux in 11 filters for each segment in each of the three detection
images.
We re-estimated the errors on the measured fluxes by random sampling of the
surrounding background. Because of drizzling and PSF matching, the noise is
correlated, so a simple pixel-to-pixel variance estimate is not sufficient.
Instead, for each of the labeled segments, we extracted the flux in 200
apertures randomly placed within a square region $\pm$ 300 pixels surrounding the
segment. For simplicity, we created circular apertures each of the same pixel
area as the segment. From these 200 segments, we computed the standard deviation
of the computed fluxes and adopted this value as the error on the originally
measured flux.
We performed this procedure for the three UV filters F225W, F275W, and F336W
only. The general S/N is much lower in the optical and IR filters, and the field
gets crowded there, making the random aperture placement impractical.

\begin{table*}[tbph]
  \caption{\label{tab:bpout_fullapert_fixmet}Best fit physical properties of
    leaker candidates}
\centering
\begin{tabular}{{@{}lcccccc@{}}}
\toprule
Filter                          & log\textsubscript{10}(M\textsubscript{\(\star\)}) & Age                      & \(\tau\)\textsubscript{del} & A\textsubscript{V}       & \(z\)                    & EW\textsubscript{Ly\(\alpha\),obs}\\
Object                          & M\textsubscript{\(\odot\)}                        & Gyr                      & Gyr                         & mag                      &                          & Å\\
\midrule
F275W-314                       & \(9.34^{+0.01}_{-0.01}\)                          & \(3.20^{+0.03}_{-0.07}\) & \(6.58^{+2.25}_{-2.35}\)    & \(0.24^{+0.02}_{-0.03}\) & \(2.00^*\)               & --\\
F275W-2055                      & \(9.10^{+0.27}_{-0.34}\)                          & \(1.07^{+0.90}_{-0.60}\) & \(5.69^{+2.96}_{-3.04}\)    & \(0.43^{+0.09}_{-0.12}\) & \(2.33^{+0.29}_{-0.29}\) & --\\
\textbf{F336W-189}              & \(9.58^{+0.01}_{-0.03}\)                          & \(0.22^{+0.01}_{-0.02}\) & \(0.13^{+0.01}_{-0.02}\)    & \(0.56^{+0.01}_{-0.01}\) & \(3.46^*\)               & 127 $\pm$ 7\\
\textbf{F336W-554}              & \(8.82^{+0.04}_{-0.04}\)                          & \(0.15^{+0.02}_{-0.02}\) & \(3.71^{+4.18}_{-2.99}\)    & \(0.82^{+0.02}_{-0.02}\) & \(2.87^*\)               & --\\
F336W-606                       & \(8.51^{+0.03}_{-0.04}\)                          & \(0.10^{+0.02}_{-0.02}\) & \(0.04^{+0.02}_{-0.01}\)    & \(0.00^{+0.02}_{-0.02}\) & \(2.73^{+0.01}_{-0.01}\) & --\\
\textbf{F336W-1013\(^\dagger\)} & \(8.50^{+0.01}_{-0.01}\)                          & \(0.08^{+0.01}_{-0.01}\) & \(4.57^{+3.73}_{-3.32}\)    & \(0.98^{+0.01}_{-0.01}\) & \(2.98^*\)               & 118 $\pm$ 30\\
\textbf{F336W-1041}             & \(9.46^{+0.03}_{-0.03}\)                          & \(1.79^{+0.06}_{-0.12}\) & \(0.48^{+0.03}_{-0.03}\)    & \(0.25^{+0.06}_{-0.04}\) & \(3.33^*\)               & 210 $\pm$ 50\\
\bottomrule
\end{tabular}
\tablefoot{
    Based on the fixed-size apertures, assuming the same
    delayed-\(\tau\)  model as before and a fiducial metallicity set to
    0.2~Z\textsubscript{\(\odot\)}. Here we also list measured
    EW\textsubscript{Ly\(\alpha\)} where available.\\
    \(^*\): Redshifts are fixed from Table~\ref{tab_redshifts}. \\
    \(^\dagger\): SFH are not well determined by our models.
}
\end{table*}



\subsection{Sorting and selection                           \label{sec:filtering}}
\label{sec:orgf3b3d63}
\subsubsection{By Signal-to-noise}
\label{sec:org90016a9}
The detection and deblending procedures described above yielded 2187, 2520, and
1804 segments in F225W, F275W, and F336W. One would probably expect more
detected objects for longer wavelength filters, but we deliberately set the
detection and segmentation limits to allow for spurious detections, of which we
expect to find more in the bluer filters where noise levels are higher and
throughput lower.

While random fluctuations can align to yield spurious detections, the
probability of this happening within the same small area of a few tens of pixels
in adjacent filters is small, and such a chance alignment would be caught when
examining the full SED. Therefore, our first selection criterion was that a
detection in any given segment be \(> 2 \sigma\) in both the detection filter
and in its red-adjacent filter, such that detections in F225W must be \(\ge\)
2\(\sigma\) also in F275W, detections in F275W must be \(\ge\) 2\(\sigma\) also
in F336W, and detections in F336W must be \(\ge\) 2\(\sigma\) also in F435W. The
dual-band SNR filtering left 1084, 1183, and 1261 segments in F225W, F275W, and
F336W, respectively.

\subsubsection{By UV colors                                  \label{sec:uvcolors}}
\label{sec:org649abfa}
As a next step, we have applied as a filtering criterion that the measured
magnitude in the detection filter be larger (fainter) than in the adjacent
filter to the red side. Typical star forming galaxies have a UV \(f_{\lambda}\)
spectrum defined as a power law with an exponent, the \emph{UV slope,} of
\(\beta \sim -2\). This corresponds to a flat spectrum in
f\textsubscript{\(\nu\)} units and a flat SED in AB magnitudes. Both dust and
ISM attenuation will affect the bluer filter more strongly than the redder one,
and thus, the observed ratio of ionizing-to-nonionizing flux will be a measure
of the escape fraction f\textsubscript{esc,LyC}. No stellar population modeling
we know of does predict any populations which intrinsically have higher
f\textsubscript{\(\nu\)} fluxes at restframe \(\sim\) 800 Å than at restframe
\(\sim\) 1000 Å, as the filters will approximately probe. Even in case they
exist, wavelength dependent attenuation by dust will suppress
F\textsubscript{800} more strongly than F\textsubscript{1100}. The only
empirically based dust attenuation law that covers the extreme UV wavelength
domain \citep{reddy2016} is particularly steep in this range and thus, even the
modest difference in rest-frame wavelength can mean a significantly stronger
attenuation at \(\lambda\)\textsubscript{rest}=800 Å than at
\(\lambda\)\textsubscript{rest}=1000 Å. In addition, LyC is absorbed when
interacting with the neutral ISM, which will suppress the
F\textsubscript{900}/F\textsubscript{1500} ratio further. LCEs have been
observed with relative (i.e., not accounting for dust attenuation) escape
fractions from almost zero \citep{leitet2011,leitet2013,puschnig2017} to close
to 100\% \citep{riverathorsen2019}. This further suppresses F900/F1500 for a
given galaxy.

Finally, at redshifts \(z \gtrsim 2\), absorption by the neutral fraction of the
Intergalactic Medium becomes significant. The average IGM attenuation due to
ionization of rest-frame \(\lambda \sim 880\) Å photons emitted at redshifts 2
(3) has been modeled by \citet{inoue2014} to be 0.3 (0.8) magnitudes,
corresponding to an average transmission factor of \(\sim\)0.76 (\(\sim\)0.48).
These numbers are however subject to a large scatter, see e.g. \citet{vasei2016}
for an example distribution at \(z=2.4\).

With these combined attenuation effects on ionizing photons emitted within \(2
\le z \le 3.5\) in mind, we require our sources to be fainter in the detection
filter than in its red-adjacent filter in order to be a believable LyC source.
Applying this criterion left 385 segments in F225W, 598 objects in F275W, and
861 objects in F336W. It is interesting to note that while we detected fewer
objects in F336W than the two other filters, more objects are left in this
filter after the SNR and color criteria have been applied.

\subsubsection{By Morphology/Visual inspection               \label{sec:morphvis}}
\label{sec:org533be7d}
After these initial filtering steps, we produced an SED plot as well as a series
of ``postage stamp'' images of all remaining objects. By visual inspection, we
removed all segments that were clearly fragments of larger nearby galaxies, as
well as a number of edge cased which were difficult to automate. In some cases,
a candidate had two nearby matches in the I17 catalog of strongly differing
redshifts, and a visual inspection was the best way to determine whether our
candidate was associated to one or the other, or the two were too strongly
blended to tell apart. Among the galaxies without matches in the I17 catalog,
some met the color selection criteria above but the combination of their
brightness, morphology and very red SED led us to reject them as nearby
interlopers. In other cases, faint detections in the UV which passed the
selection criteria, showed so strong blue colors that they were undetected in
the optical and IR filters and thus not suitable for further investigation. A
few spurious detections had made it this far as well but was easily identified
from inspecting their SED and images.

\subsubsection{By spectroscopic redshifts}
\label{sec:org0d21d3b}

Spectroscopy, where available, is the most reliable way of measuring redshifts.
\citet{inami2017} published a catalog of 1574 spectroscopic redshifts in the HUDF,
of which 1338 are reported as high quality redshifts. We matched our candidate
objects with the I17 catalog by searching for sources within a radius of \(0
\farcs 75\), slightly above the reported FWHM seeing in \citet{inami2017}, beyond
which contamination from nearby compact sources is weak, and listed for each of
our sources all MUSE sources within this radius. In order to check for
contamination and look for sources not found in the catalog, we retrieved the
publicly available MUSE-DEEP science cubes. For each galaxy in the sample
remaining from the filtering process in sect. \ref{sec:filtering}, we extracted a
spectrum from the muse cubes in the following way:

For each object, we determined whether it was contained in the footprint of each
of the MUSE-DEEP cubes. For objects contained in a given cube, we extracted all
pixels inside a circular mask of radius \(0 \farcs 75\) and, for each velocity
slice, summed up the flux inside this aperture. Objects found inside the
footprints of multiple cubes were median-stacked. We then inspected the spectra
for line emission; if multiple lines were present, their wavelength separation
reveals which lines we were looking at. If only one line was present, we
preliminarily assumed it to be Ly\(\alpha\). Spectroscopic redshifts found by this
procedure are, where available, tabulated for objects of interest in Table
\ref{tab_redshifts}.

Since our extraction was less sophisticated than that of \citeauthor{inami2017},
the resulting spectra are not of the same quality as in I17. Indeed, in a number
of cases, where those authors' redshift determination depends on absorption
features, we did not have sufficient SNR to discern individual absorption
features. In such cases, we have listed the reported values and compared them to
available photometric redshifts and, where not consistently contradicted by
these, we have adopted the redshifts of MUSE-DEEP for the given object.

\subsubsection{By photometric redshifts}
\label{sec:org0d9894d}

Included in the catalog of \citet{rafelski15_uvudf} are photometric redshifts
computed in two different ways, the \emph{Bayesian photometric redshift}
\citep[BPZ,][]{benitez2000,benitez2004,coe2006}, and the redshift as computed by
the software package EAZY \citep{brammer2008eazy}. In addition, in this work, we
computed photometric redshifts for the subset of our detected objects that made
it through the elimination process in section \ref{sec:filtering}, using the
stellar population modeling package \texttt{BAGPIPES}
\citep{carnall2018}\footnote{\texttt{BAGPIPES} uses the \citet{bruzual2003}
  stellar models, the \texttt{Multinest} \citep{feroz2019} sampling algorithm
  and its python interface \texttt{PyMultiNest} \citep{buchner2014}.}, and the
photometric fluxes of the full set of 11 filters where available. We assumed a
single population with a \emph{delayed \(\tau\) model}, defined as:
SFR\((t) = t \times e^{-t/\tau}\), where \(\tau\) is the timescale of
exponential decay of star formation. This model provides a good balance between
flexibility and simplicity, with exponential drop-off allowing a span from the
strongly bursty to much more extended SFHs. Of the dust attenuation laws
implemented in \texttt{BAGPIPES}, we assumed \citep{calzetti2000} law,
extrapolated to the FUV by a power law \citep{carnall2018}, allowing \(A_V\) to
vary between 0 and 4 mag, and allowed the redshift to vary freely between 0 and
10. We enabled \texttt{BAGPIPES}' built-in support for \texttt{CLOUDY}
\citep{ferland2017} assuming \(\log U = -3\) and otherwise using the default
settings; this would allow \texttt{BAGPIPES} to also correctly model strong
emission line galaxies. \texttt{BAGPIPES} assumes pure Case B recombination and
thus zero ionizing escape from the model galaxy \citep[][Sect.~3]{carnall2018}.

For IGM absorption, \texttt{BAGPIPES} adopts the analytical version of the IGM
transmission law for \(T(\lambda, z_obs)\) by \citet{inoue2014}, assuming that
all flux with \(\lambda < 912Å\) is absorbed \citep[this is however already the
case when nebular emission is enabled, see][Sect.~3]{carnall2018}. We keep in
mind that the assumption of zero transmission of IGM could lead
\texttt{BAGPIPES} to interpret all data points with nonzero flux as being
non-ionizing and thus be skewed towards lower-redshift solutions. Mostly, the UV
filters have low SNR compared to the optical and IR filters, but if the
optical/IR data is more or less evenly consistent with two different models, the
presence of nonzero UV fluxes may tip it in favor of the lower-redshift
solution.

We then compared the photometric redshifts to those published by
\cite{rafelski15_uvudf}, as well as the MUSE spectroscopic redshifts published
by \cite{inami2017}. Where spectroscopic redshifts exist, we have relied on
those. Due to the assumption hard coded into \texttt{BAGPIPES} that no ionizing
flux reaches the telescope, the minimization algorithm will, if allowed, attempt
to adjust the redshift to interpret all bands with measured flux to fall below
the Lyman edge at \(\lambda\)\textsubscript{rest}=912 Å. However, the UV bands
have considerably poorer SNR and are less numerous than the optical bands, which
counteracts this effect. See discussion of individual objects in
sect.~\ref{sec:org5312824}.

\subsubsection{Test of redshift with emission lines} \label{sec:org531ed8e} For
the galaxies in the final candidate selection which show Lyman-\(\alpha\)
emission, we have measured the equivalent width by measuring the
continuum-subtracted line flux in the MUSE spectra, and dividing it by the flux
density as measured in an aperture of the same size in the \emph{HST} filter
containing \(\lambda\)\textsubscript{obs,Ly\(\alpha\)}. The continuum in MUSE is
noisy, and an unknown fraction of it consists of scattered light from the
atmosphere, so we preferred the deeper, less contaminated continuum levels from
HST imaging. In the case of F336W-1013, Ly\(\alpha\) falls between F435W and
F606W, and we have used a simple average of the fluxes in those filters to
compute the equivalent width. The resulting observed-frame EWs are tabulated in
table \ref{tab:bpout_fullapert_fixmet} along with a number of physical
parameters derived from a second round of SED modelling using \texttt{BAGPIPES},
see Sect.~\ref{sec:faphot}.

The computed equivalent widths are useful to check for misclassified spectral
emission lines. The observed-frame EWs range from 118 to 210 Å which in the
restframe yield \(\sim\) 30--50Å. Other comparably strong emission lines are
found from rest-frame [\ion{O}{ii}] \(\lambda \lambda\) 3727,3929 Å and redward,
where the redshift corrections to the measured EWs are lower by roughly a factor
3 or more. This leaves only few lines which could be consistent with the
measured values. For example, [\ion{O}{ii}] \(\lambda \lambda\) 3726,3729, could
possibly be consistent with the measured lines, but this feature is a multiplet;
and at such redshifts, lines such as H\(\beta\), the [\ion{O}{iii}] doublet at
\(\lambda \lambda\) 4960,5008 Å, and H\(\alpha\) would all fall within the
redshift range of the spectrograph, none of which are found. We also note the
dual asymmetric peaks shape of the line in F336W-189, a typical and unique
feature of Ly\(\alpha\); and the clear asymmetric shape of F336W-1041, typical
of a Ly\(\alpha\) line at modest SNR.

\subsubsection{Final selection}\label{sec:org5cf08e5}

The previous steps in combination left us with a selection of seven objects of
interest, listed with their R15 and I17 counterparts where available in
Table~\ref{tab:baseprops}. The strongest candidates shown in \textbf{bold} type
in Table~\ref{tab:baseprops} and subsequent tables. These all have spectroscopic
redshifts in I17 at reported at confidence levels 2 (F336W-1013, F336W-1041) or
3 (F336W-189, F336W-554), and we were able to confirm the spectroscopic redshifts
from our own extracted spectra. In addition, these candidates all have flux
measured at at least 2\(\sigma\) in at least two filters which are entirely in
the rest-frame Lyman Continuum at the spectroscopic redshift, except for
(F336W-1013).

In addition, we have identified three second-tier candidate objects which we
consider highly likely, but not definitive, leakers. Two of them have no
spectroscopic redshift, but the available photometric redshifts are high enough
that the flux in the FUV filters must be dominated by LyC; and one has a
convincing spectroscopic redshift which is however only marginally consistent
with being a LyC leaker. See Sec.~\ref{sec:org5312824} for a discussion of each.

\begin{figure*}[tbhp]
  \centering
  \includegraphics[width=.9\textwidth]{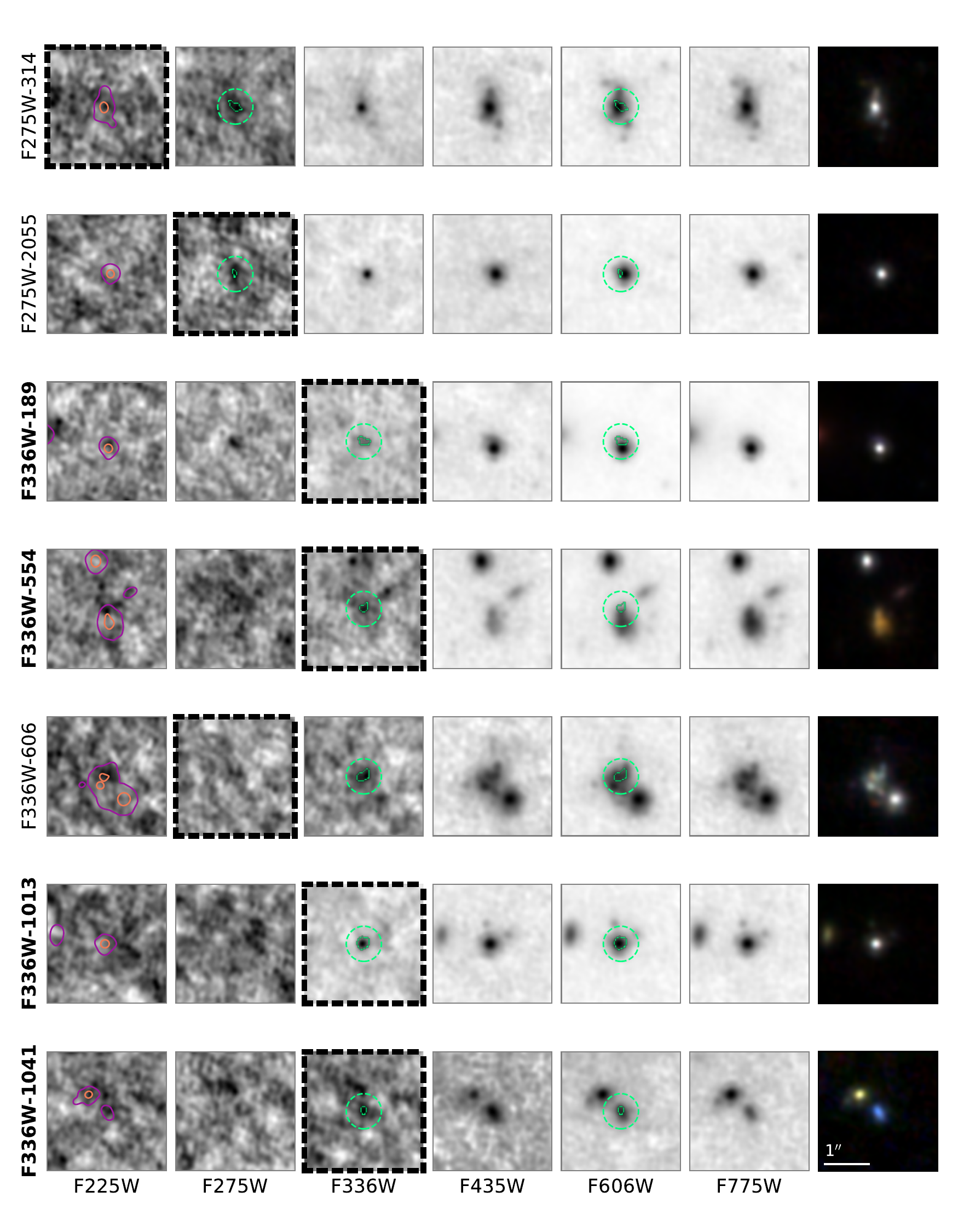}
  \caption{Postage stamp images in the filters F225W--F775W of the candidate
  objects, as well as a composite image of F435W, F606W and F775W for each
  candidate. The detection segment and the larger photometric aperture are shown
  in green in F606W as well as in the detection filter for each candidate. The 
  reddest LyC-dominated filter is framed in thick black dashes for each 
  candidate, and contours of the brightness in F775W are shown in F225W to 
  provide a morphological comparison between the rest frame far- and near-UV 
  bands. The cut-outs are $2.7\arcsec$ on a side.\label{fig:stamps}}
\end{figure*}

Fig.~\ref{fig:stamps} shows $1.7\arcsec$ square cut-outs in the filters from
F225W to F775W, as well as an RGB composite of the filters F775W, F606W, and
F435W in the R, G and B channels, respectively.

\section{Candidate characterization                        \label{sec:candchar}}
\label{sec:org23193da}
\subsection{Fixed-aperture photometry \& global SED modeling \label{sec:faphot}}
\label{sec:org6a09af0}

Up to this point, all fluxes have been measured in the segments created from the
source detection in the UV filters. The young stellar populations that dominate
production of UV photons is generally significantly more compact than the older
population which often dominates the optical and IR bands. Our choice of
aperture thus yields comparatively high fluxes in the UV filters compared to
what would result from photometry of the entire galaxy.

\begin{figure}[tbhp]
  \centering
  \includegraphics[width=\columnwidth]{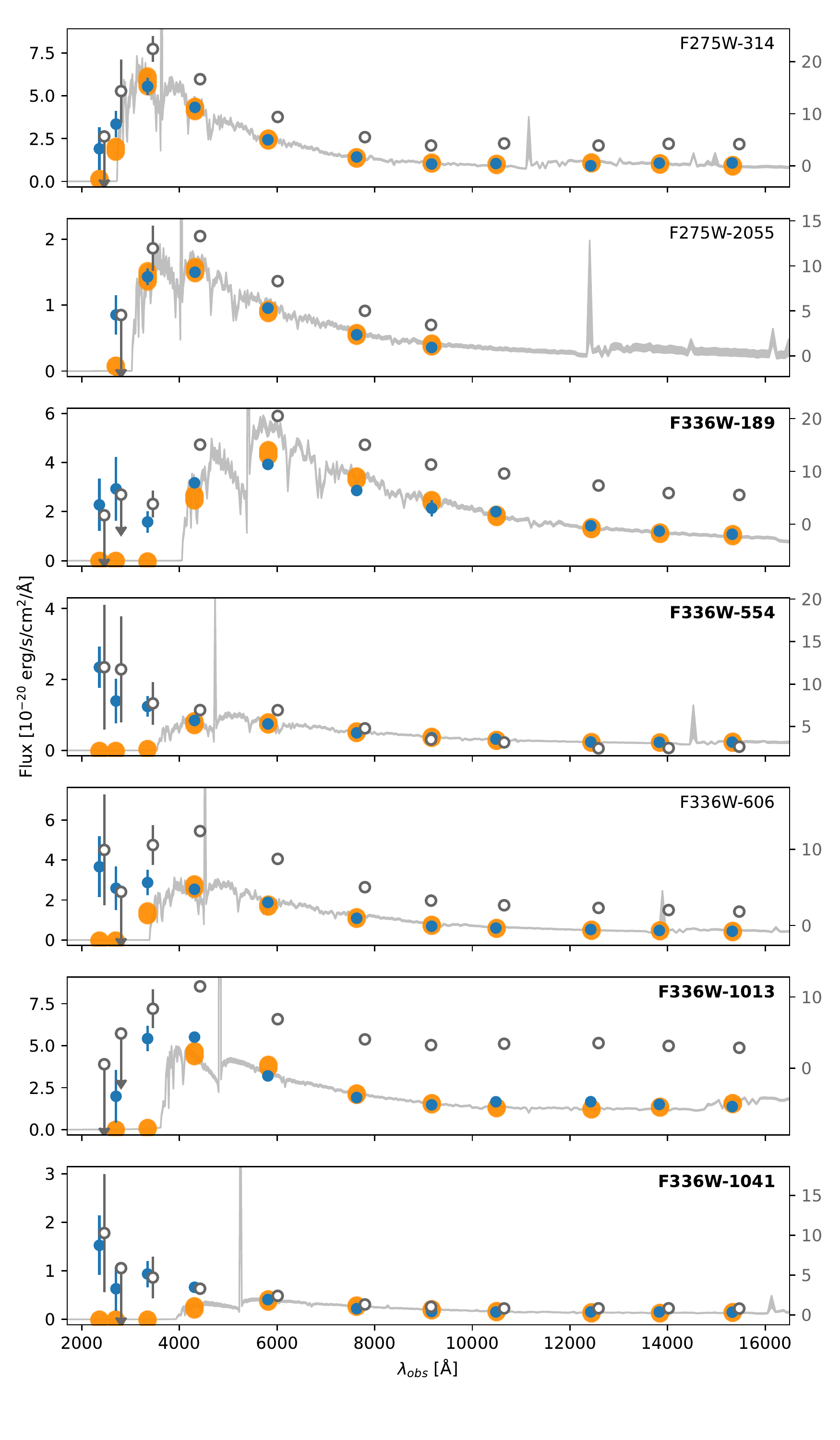}
  \caption{\label{fig:sedfig}
  Segment fluxes (blue) and fixed-aperture fluxes (empty circles with the flux 
  scale on the right) shown with \texttt{BAGPIPES} posterior photometry 
  (orange dots) and model spectra (gray shading).}%
\end{figure}

To derive global physical properties for the candidate galaxies rather than
local properties for the often small regions defined by the segmentation maps,
we have supplemented these measurements with fixed-aperture photometry for the
objects we found to be LyC leaker candidates. We chose a fixed circular aperture
\(0\farcs8\) in diameter, which is large enough to encompass the majority of
most galaxies in the relevant redshift range, and small enough to keep the risk
of contamination from nearby objects small. In fig.~\ref{fig:leakerSEDs}, we
show the resulting spectral energy distributions for the candidate objects
listed in Table~\ref{tab_redshifts}, this time shown in AB magnitudes. The same
magnitudes are tabulated in table~\ref{tab:fluxes}. Note that that this larger
aperture made the MC sampling of uncertainties unfeasible. With aperture size
grows also the probability of catching a bright nearby object in the aperture
which will skew the standard deviation. For this reason, we have found this
method unfeasible with a \(0\farcs8\) aperture, and have opted to simply report
the uncertainties found from the RMS frames. Additionally, the compactness in
the UV bands has the consequence that a larger number of pixels will contribute.
As a result, the errors reported from fixed apertures are markedly larger than
from segmentation images, with some measurements being consistent with 0 within
1\(\sigma\). We have opted to report these as measurements with error bars
rather than upper limits in table~\ref{tab:fluxes} and
figures~\ref{fig:leakerSEDs} and \ref{fig:sedfig}, but have not based any
quantitative conclusions on these.

After completing the redshift and color selection process, we revisited the
\texttt{BAGPIPES} SED fitting for the remaining candidate objects, this time
based on fluxes extracted from the fixed-size apertures. This time, we fixed the
redshifts at the spectroscopic values where available, and in the two cases
where this was not the case, we left it open as a free parameter. Again, we
assumed a delayed \(\tau\) model, finding it to be the one single model which
can cover the widest range of physical properties, but this time we imposed a
metallicity constraint of \(Z \le Z_{\odot}\). In
table~\ref{tab:bpout_fullapert_fixmet}, we tabulate the best-fit values of core
properties obtained from this second run.

\subsection{Ionizing escape fractions}
\label{sec:org763ea23}\label{sec:fescape}

Following \citet{steidel2001,siana2007,siana2010,grazian2016,fletcher2018} we
define the \emph{relative} and \emph{absolute} escape fraction in the following
way: 

\begin{align}
	f_{\text{esc}}^{\text{rel}} &= \frac{(f_{1500}/f_{900})_{\text{Int}}}{(f_{1500}/f_{900})_{\text{Obs}}} \times\ \frac{1}{T_{\text{IGM}}}, \\
        f_{\text{esc}}^{\text{abs}} &= f_{\text{esc}}^{\text{rel}} \times 10^{-0.4 \times A_{1500}},
\end{align}

\noindent where $(f_{1500}/f_{900})_{\text{Int}}$ and
$(f_{1500}/f_{900})_{\text{Obs}}$ are the intrinsic and observed ratios of flux
density at rest-frame wavelengths 1500 and 900 Å, $T_{\text{IGM}}$ is the IGM
transmission coefficient along a given line of sight, and $A_{1500}$ is the
attenuation at rest-frame wavelength 1500 Å in magnitudes. Often, an intrinsic
flux ratio of $(f_{1500}/f_{900})_{\text{Int}} = 3$ (in $f_\nu$ units) is
assumed \citep[e.g.][]{steidel2001,grazian2016,grazian2017}; however, the value
has been found to vary from 1.7 at a burst age of 1 Myr, to 7.1 at 0.2 Gyr of
age, using the stellar population library of
\citet{bruzual2003}\citep{grazian2016}, and where possible, it is preferable to
estimate the intrinsic ratio for each individual galaxy. 

In this work, we have opted to find $f_{\text{esc}}^{\text{abs}}\times
T_{\text{IGM}}$ directly from the \texttt{BAGPIPES} models, and then applying
the dust correction described above to find $f_{\text{esc}}^{\text{rel}}\times
T_{\text{IGM}}$, and then apply a corrective term to account for IGM absorption
as described in the next section.

We have estimated the absolute ionizing escape fraction by comparing observed
LyC fluxes to modeled intrinsic fluxes in the following way: For each best-fit
\texttt{BAGPIPES} model, we constructed a model galaxy based on the fit, but
omitting the effects of internal dust, ISM, and IGM absorption. We then compared
the observed LyC fluxes in the reference LyC filters shown in
Fig.~\ref{fig:stamps}. The fraction of model-to-observed flux is the combined
transmission galaxy ISM and the IGM. We then estimated the \emph{relative}
escape fraction of each candidate, by applying the best-fit dust attenuation at
rest-frame 1500 to the absolute fraction.

\begin{figure*}[tbhp]
  \centering
  \includegraphics[width=.75\textwidth]{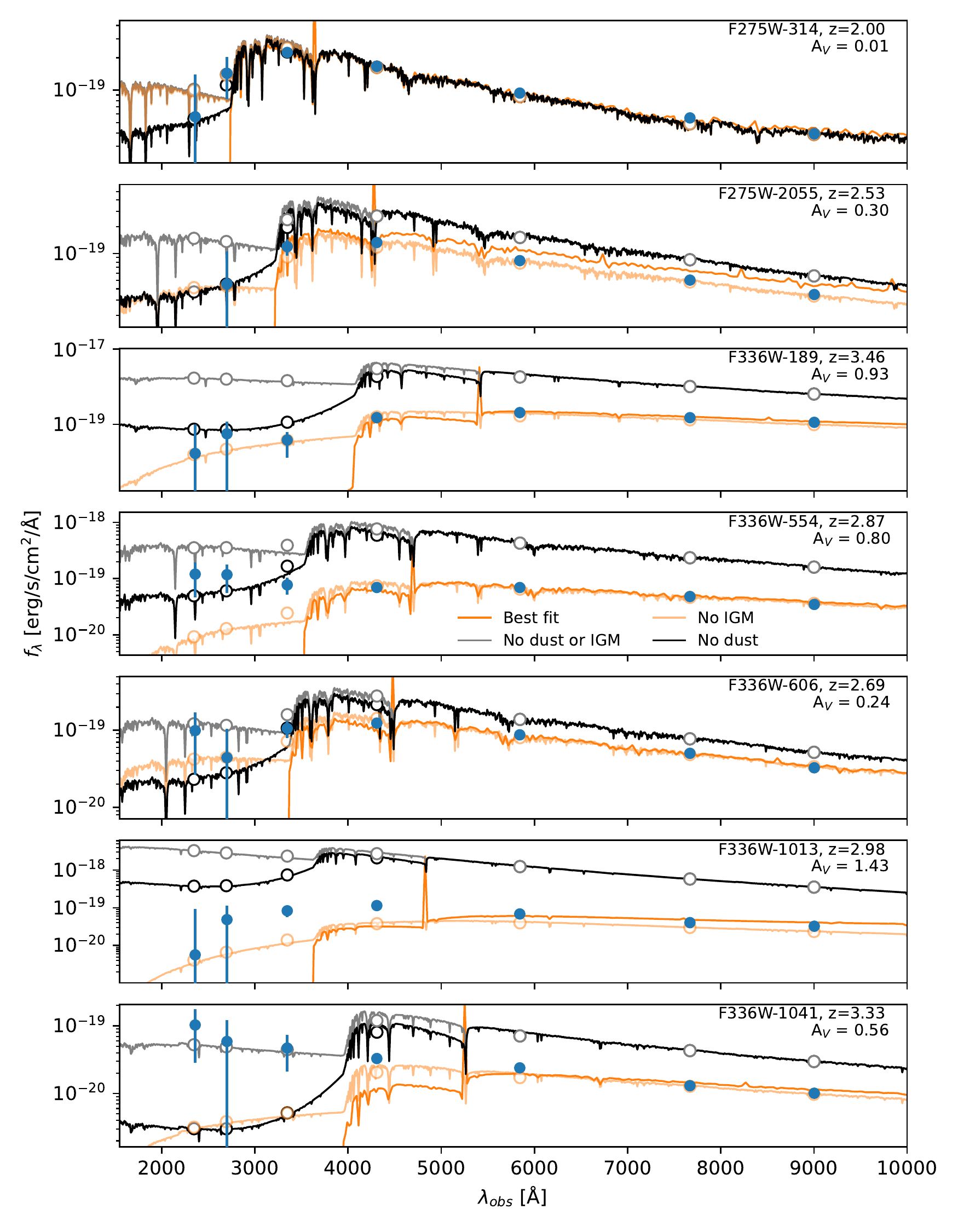}
  \caption{Dust-, ISM-, and IGM-corrected model spectra of the candidates, as
  well as observed (filled circles) and model predicted (open circles). Orange 
  spectra are the \texttt{BAGPIPES} best fit to the observed data points. Model
  spectra generated from the same best-fit parameters, but with ISM and dust
  removed, are shown in black. Gray spectra show the ISM- and dust-free model
  spectra corrected for the \texttt{BAGPIPES} assigned average redshift
  dependent IGM absorption.\label{fig:corrections}}
\end{figure*}

Fig.~\ref{fig:corrections} shows for each candidate the observed fluxes (blue,
filled circles), along with the best-fit, IGM-attenuated spectrum (strong
orange) and the dust-and ISM-free, but still IGM-attenuated, model spectrum
(black). Each of these is additionally shown with the
\texttt{BAGPIPES}-generated IGM attenuation removed, shown in a lighter tone
(pale orange, gray). Synthetic HST filter photometry, also generated by
\texttt{BAGPIPES}, is shown as open circles in the colors of the corresponding
model spectrum. The absolute escape fraction can then be visually understood as
the fraction of the values marked by the filled blue to the open gray circle in
the reference LyC filter. It is worth noting that while a similar interpretation
of the relative escape fraction as ratio of the filled blue to open light orange
circles would be physically correct, that is not the value found as
$f_{\text{esc}}^{\text{rel}}$ using the convention described above. The latter
convention assumes the dust attenuation to be the same at 900 and 1500 Å and is
thus lower than what would be found by directly measuring the ratio of escaping
to dust-corrected intrinsic flux at 900 Å; the discrepancy between the two will
depend on the choice of dust attenuation law.


It is interesting to note from inspection of Fig.~\ref{fig:corrections} that
while the relative escape fractions can be even quite dramatically higher than
100\%, all \emph{absolute} escape fractions seem roughly consistent with being
$\lesssim 1$, even for an object like F336W-1013, for which the \emph{relative}
escape fraction is $\sim 450\%$ (see Table~\ref{tab:fescs}) after IGM correction.
The relative escape fraction is computed based on the assumption that the light
from the stars responsible for emitting LyC undergo the same dust attenuation as
the full stellar population, on which the galaxy model is built. However,
recent studies of the Magellanic clouds
\citep{ramachandran2018,ramachandran2018b,ramachandran2019,doran2013} have shown
that the ionizing output from a galaxy can be completely dominated by as little
as one to a few dozen extremely luminous, massive stars.  If these stars are
seen along a privileged line of sight through the dust in their galaxy, they
could conceivably be much less attenuated than the greater stellar population.
This could either be due to a highly clumpy dust geometry, or to the stars
dominating LyC emission being located away from the stars emitting the bulk of
rest-frame 1500Å photons \citep[see e.g.][for an example of O stars residing in
the tidal bridge between the Magellanic Clouds]{ramachandran2021}. 

One candidate, F336W-1041, has an absolute escape fraction of $\sim 100\%$ even
before IGM correction and thus requires an essentially completely dust- and gas
free line of sight between the emitting stars and the telescope for the observed
escape fraction to be physical, and other candidates require similar,
intrinsically quite unlikely, constraints on intervening dust and HI column to
be fulfilled. However, it should be kept in mind that we have inferred a strong
selection bias for exactly such conditions. 

\subsection{IGM correction}

\texttt{BAGPIPES} accounts for IGM attenuation following \citet{inoue2014},
applying an average redshift-dependent IGM attenuation curves derived in that
paper \citep{carnall2018}. While this can be very useful to generate a
realistic, generic model galaxy at a given redshift, IGM absorption along a
given line of sight is highly stochastic, as shown in e.g.
\citet{vanzella2015,vasei2016,bassett2021,bassett2022}. Specifically,
\citet[][Fig. 6]{jones2021} show that the average IGM optical depth to LyC
changes drastically over the range of wavelengths of this work.  In addition, as
discussed above, in this work, we are selecting specifically for
high-transmission lines of sight, and thus an unbiased average is unlikely to be
a good approximation to lines of sight actually observed. 

\begin{table*}[tbph]
  \caption{\label{tab:fescs}Escape fractions and IGM correction}
\centering
\begin{tabular}{{@{}lrrrrr@{}}}
\toprule
Object                          & \(f_{\text{esc}}^{\text{abs}}\times T_{\text{IGM}}\)\tablefootmark{a} & \(f_{\text{esc}}^{\text{rel}}\times T_{\text{IGM}}\)\tablefootmark{b} & $T_{\text{IGM}}$   &\(f_{\text{esc}}^{\text{rel}}\)& $f_{\text{esc}}^{\text{abs}}$ \tablefootmark{c}\\
\midrule
F275W-314           & \( 51 \pm 33   \%\) & \( 57\%\) & \(72  \pm  9\%\) & \(82  \pm 12\%\)  & \( 77 \pm  9 \%\) \\
F275W-2055          & \( 57 \pm 19   \%\) & \(105\%\) & \(66  \pm  8\%\) & \(153 \pm 14\%\)  & \( 76 \pm  9 \%\) \\
\textbf{F336W-189}  & \(  3 \pm  0.8 \%\) & \( 23\%\) & \( 7  \pm  5\%\) & \(325 \pm 200\%\) & \( 36 \pm 22 \%\) \\
\textbf{F336W-554}  & \( 28 \pm  7   \%\) & \(203\%\) & \(49  \pm 11\%\) & \(441 \pm 110\%\) & \( 67 \pm 15 \%\) \\
F336W-606           & \( 40 \pm 17   \%\) & \( 67\%\) & \(42  \pm  3\%\) & \(146 \pm 12\%\)  & \( 91 \pm  6 \%\) \\
\textbf{F336W-1013} & \(  4 \pm  0.6 \%\) & \(103\%\) & \(24  \pm 15\%\) & \(442 \pm 295\%\) & \( 15 \pm 10 \%\) \\
\textbf{F336W-1041} & \(108 \pm 30   \%\) & \(397\%\) & \(100 \pm  -\%\) & \(397 \pm -\%\)   & \(100 \pm  - \%\) \\
\bottomrule
\end{tabular}
\tablefoot{
        \tablefoottext{a}{
                The quoted uncertainties here are an estimate based on the
                assumption that the true SNR of the large-aperture photometry is
                the same as the corrected SNR in the segmentation map photometry
                (see Sect.~\ref{sec:mcerrs}), and assuming no uncertainties due
                to the stellar population modeling. They are not propagated further.
        }\\
        \tablefoottext{b}{
                No uncertainties are quoted here, as the relative and absolute
                escape fraction only differ by a dust correction term adopted
                from the \texttt{BAGPIPES} best-fit galaxy model.
        }\\
        \tablefoottext{c}{
                Uncertainties quoted here are based solely on the simulated IGM
                transmission distribution. Uncertainties in this quantity and
                in $f_{\text{esc}}$ are highly degenerate.
        }
}
\end{table*}

To obtain a realistic estimate of the impact of IGM absorption on the observed
LyC fluxes, we first run Monte Carly simulations of 1000 random, unbiased lines
of sight following the prescription given by \citet{inoue2014}.
We next imposed as a prior constraint on these lines of sight that they must be
consistent with an $f_{\text{esc}}^{\text{abs}} \leq 100\%$. All lines of sight
not meeting this requirement were removed from the LOS distribution. Next,
$f_{\text{esc}}^{\text{abs}}$ was found assuming the IGM opacity of each line of
sight, and a resulting distribution of escape fractions was found. For each
candidate, we then report the mean and standard deviation of this distribution
as the escape fraction. One galaxy, F336W-1041, has $f_{\text{esc}}^{\text{abs}}
\approx 100\%$, requiring an IGM transmission of unity, to which we have fixed
it. 

In Table~\ref{tab:fescs}, we report the measured and absolute escape fractions
without IGM correction, as well as the derived mean and standard deviation of
the IGM transmission coefficient $T_{\text{IGM}}$, and the resulting
IGM-corrected absolute escape fractions. Uncertainties quoted for the latter are
based solely on the IGM distribution; the galaxy models and measured LyC fluxes
are assumed as is, without uncertainties. The uncertainties on the model
galaxies are dominated by systematics due to the choice of star formation
history, dust attenuation law etc., and the uncertainties on the IGM
transmission and the measured LyC fluxes are highly degenerate. A stricter
treatment of uncertainties of the escape fractions would entail a full MC
sampling of the LyC fluxes for a variety of SF histories and dust attenuation
models, as well as creating a full $T_{\text{IGM}}$ distribution for each
realization, which is outside the scope of this work.

\subsection{Contribution to ionizing background} 
\label{sec:org9e2158a} 

We have estimated the contribution from our candidate objects to the
metagalactic ionizing background at redshifts \(2 < z < 3.5\) broadly following
the methodology of \citet[][Sect. 6]{jones2021}. We have converted the measured
flux densities to an ionizing emissivity at 900 Å $\epsilon_{900}$, defined as
the luminosity density at 900 Å per unit frequency per unit comoving volume.

\begin{figure*}[tbh]
\centering
\includegraphics[width=0.8\textwidth]{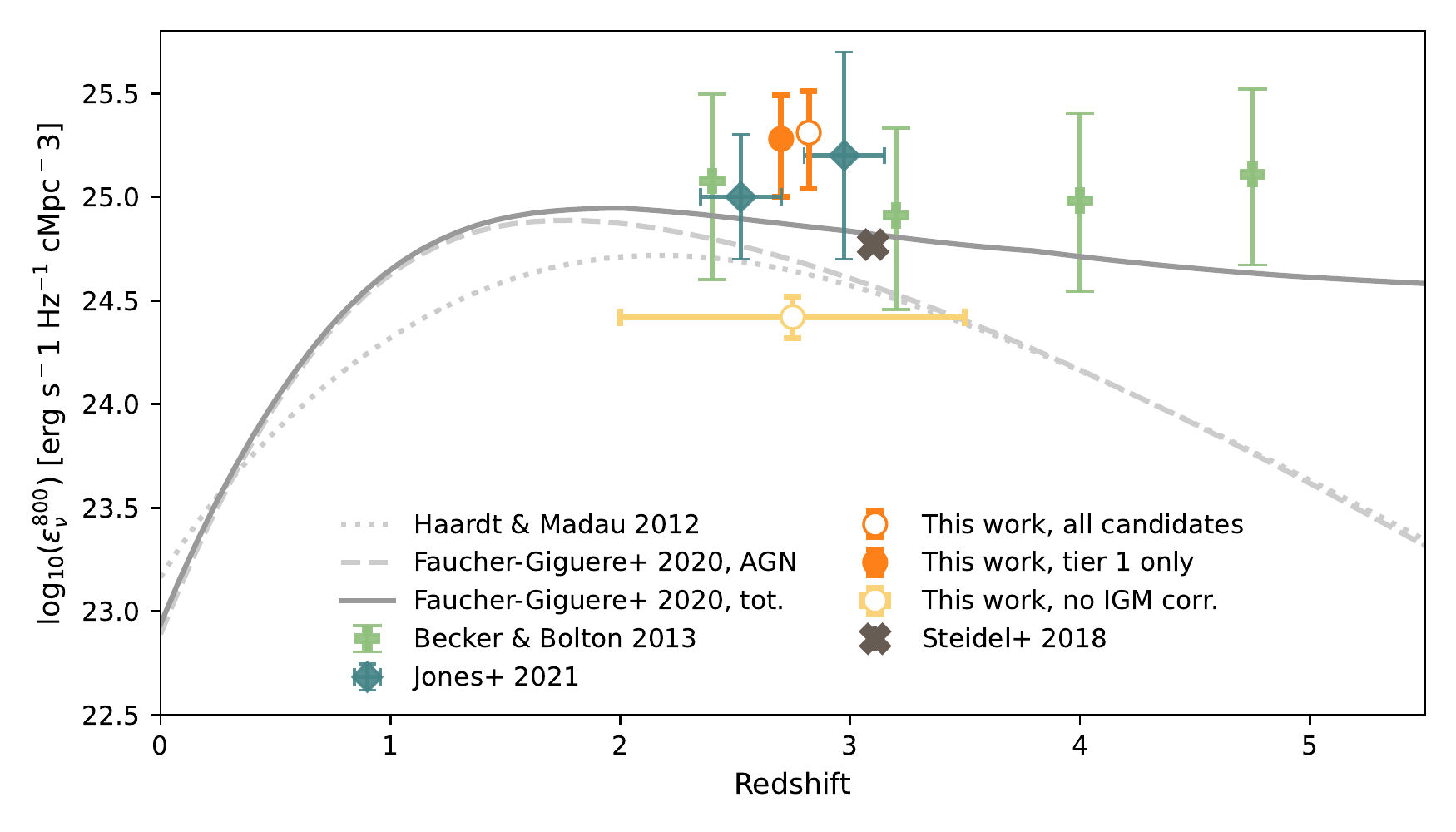}
\caption{\label{fig:epsilons}Ionizing volume emissivity derived from the
  galaxies of this work compared to other measurements and predictions in the
  literature. Gray dashed and dotted lines show the theoretical models for
  bright AGN by \citet{fauchergiguere2020} and \citet{haardt2012}, and the fully
  drawn line shows the combined model of AGN and star formation from
  \citet{fauchergiguere2020}. Light green plus signs show the ionizing field in
  various redshift bins measured from the IGM properties by
  \citet{becker.bolton2013}. Dark green diamonds show the emissivity from star
  formation measured by \citet{jones2021}, and the single dark cross shows the
  estimate from \citet{steidel2018}. The filled orange marker shows the value
  derived from only the 4 tier 1 candidates in this work, while the open orange
  marker shows the value derived assuming that all the candidates are bona fide
  leakers. The open yellow marker shows the same as the open orange marker, but
  not corrected for IGM absorption. The orange and yellow markers are showed a
  bit offset on the redshift axis for visibility, but all are measured in one
  single redshift bin of \(2 < z < 3.5\) (illustrated with the horizontal yellow
  bar), with an average redshift of 2.81.}
\end{figure*}

For each of the candidates, we have used the flux density in the reference LyC
filter shown in Fig.~\ref{fig:stamps}, and assumed the measured filter average
as the value at 900 Å. This conversion assumes that the measured flux in this
filter is a good representation of the spectral flux density at rest-frame 900 Å
or, equivalently, that the spectral shape of the Lyman Continuum in the
wavelength ranges covered by these filters is constant in $f_\nu$ units. Looking
at the fluxes in Fig.~\ref{fig:sedfig}, the measured differences between the
reddest and neighboring bluer filter are nowhere more than $\times \sim 1.5$,
making this an imperfect but reasonable assumption. 

We computed an ionizing luminosity density from each measured flux density by
correcting it for IGM absorption using the values derived in the previous
section, and multiplying the result by \(4 \pi D_L^2\), where \(D_L\) is the
luminosity distance at the given redshift. We then computed the comoving volume
defined by the \(d_A = 160''\) angular size of the UVUDF and the redshift range
defined by our filters, \(2 < z < 3.5\), as:

\begin{align}
V_c &= \int_{2.0}^{3.5} A_c(d_A, z) \frac{dr_c}{dz} dz \\
   &= 30100\, \mathrm{cMpc}^3,
\end{align}

\noindent{}where \(A_c(z)\) is the comoving area defined by the angular size
\(d_A\) of the field at redshift \(z\), and \(dr_c(z)/dz\) is the thickness of
the slab defined by the redshift difference \(dz\) at redshift \(z\). 
We have assumed that measured LyC emission is isotropic which is not generally a
good approximation for single galaxies, but which from symmetry considerations
should be a reasonable assumption for the number of sources existing in the
HUDF within our redshift range. We have found found a resulting value of
\(\log_{10}(\epsilon) = 25.32^{+0.25}_{-0.21}\) \((25.29^{+0.27}_{-0.22})\) erg
s\textsuperscript{-1} Hz\textsuperscript{-1} cMpc\textsuperscript{-3} for the
full (tier 1 only) set of candidates.

This value is shown in figure~\ref{fig:epsilons} along with predictions and
measurements from the literature. We find that our values are consistent with
those found by \citet{jones2021}. Once corrected for IGM attenuation, the
resulting emissivity from star forming galaxies is sufficient to maintain the
measured ionization state of the Universe up to redshift \(\sim\) 3.5. The value
we find for $\epsilon_{900}$ is rather high, but still consistent with, values
previously found by e.g.~\citet{jones2021}, and is dominated by one galaxy,
F336W-189, which contributes 83\% of the total, IGM-corrected ionizing
luminosity, owing at least in part to a strong IGM correction. With its high
redshift and low value of $f_{\text{esc}^{\text{abs}}} \times T_{\text{IGM}}$,
it allows allows for low-transmission IGM lines of sight and thus drives the
median IGM correction up.

We estimated the effect of cosmic variance on our results using the
\texttt{QUICKCV} cosmic variance calculator \citep{newman2014,moster2011} and,
independently the Trenti \& Stiavelli online cosmic variance
calculator\footnote{\url{https://www.ph.unimelb.edu.au/\~mtrenti/cvc/CosmicVariance.html}}
\citep{trenti2008}. Both yielded an estimated cosmic variance of \(\sim\)15-20\%
on the emissivity, significantly less important than the uncertainties stemming
from photometry and the inherent stochasticity of the IGM.

\section{Discussion                                       \label{sec:discuss}}
\label{sec:org3a19d10} Resulting from our selection and inspection criteria, we
have compiled a set of candidate LyC leaker galaxies falling into two tiers as
described in Sect.~\ref{sec:org5cf08e5}. Based on best-fit models of their
stellar populations, we have derived intrinsic and dust-attenuated LyC outputs
and, based on these as well as Monte Carlo simulations of the IGM, derived the
absolute and relative LyC escape fraction of each candidate galaxy. We have
found that while some galaxies have surprisingly strong LyC fluxes, they are all
roughly consistent with an \emph{absolute} escape fraction of $\lesssim 100\%$.
However, some of the candidates show \emph{relative} escape fractions of several
hundred percent, something which requires the LyC emitting stars to be seen
through a considerably lower dust column than the general stellar population
dominating the optical and IR filters. We discuss these cases below.

We then used the resulting IGM absorption distributions to convert the observed
LyC flux densities to a cosmic volume emissivity $\epsilon_{900}$, and compared
it to theoretical and observed values in the literature. We find a value which
is surprisingly high, but consistent with measured values from the literature
\citep{becker.bolton2013,jones2021}. The emissivity is dominated by one galaxy,
F336W-189, which contributes 82\% of the combined volume emissivity, with
F336W-1013 contributing an additional 10\%. Both galaxies have high redshifts
and as a result, a $T_{\text{IGM}}$ -distribution skewed towards low
transmission and thus a strong ISM correction. Despite the fact that as much as
four of our candidate galaxies are unlikely to have been found in any typical
preselection methods, it is thus not immediately clear how important their
contribution to the cosmic ionizing background is, mainly due to the statistical
uncertainty in the IGM corrections.

\subsection{Strongest candidates}
\label{sec:org527a5ae}

\begin{description} \item[{\textbf{F336W-189:}}] This galaxy was also identified
        by \cite{saxena2021}, and is our strongest candidate. It is bright, has
        a strong and unmistakable Lyman-\(\alpha\) emission line (see
        fig.~\ref{fig:OOIspecs}) which determines its redshift beyond any
        reasonable doubt, and the MUSE and \texttt{Photutils} centroids are
        coincident to within \(0\farcs1\). It is detected in three bands in LyC,
        and its spectral energy distribution is textbook example of what we
        would expect a star-forming galaxy with a modest
        \(f_{\mathrm{esc}}^{\mathrm{LyC}}\) to look like. It was detected in
        F275W but rejected because it is fainter in F336W than F275W, because
        both filters are in the rest-frame LyC. We find a stellar mass of
        \(\log(M_*)\) = 9.58 $\pm$ 0.02, the most massive of the candidates. The
        absolute escape fraction is 36\% $\pm$ 22\% and, after dust correction,
        we find a relative escape fraction of 325\%. The Ly\(\alpha\) line
        profile is double-peaked with a peak separation of \(\Delta v \approx
        620\) km/s. \citet{verhamme2015} found that a peak separation of
        \(\Delta v \lesssim 300\) km/s corresponds approximately to \(\log
        N_{HI} \approx 17\), at which point \(\tau_{900Å} \approx 1\), meaning
        that $f_{\text{esc}}^{\text{rel}}$ (LyC) for this galaxy is a good deal
        higher than expected from the peak separation (but as discussed above,
        the clumpy nature of dust may have inflated the relative escape fraction
        substantially and makes a comparison to theoretical values tricky). This
        galaxy also dominates the found value of $\epsilon_{900}$, contributing
        83\% of the total IGM-corrected ionizing flux in the searched volume.
        This large contribution owes at least in part to the low measured,
        IGM-uncorrected escape fraction, which is consistent with IGM lines of
        sight 

\item[{\textbf{F336W-554:}}] This galaxy does not show any line emission,
        emission in Ly\(\alpha\), but its spectroscopic redshift of 2.87 from
        I17 is reported with a confidence level of 3, the highest in that work,
        based on multiple absorption features. The two photometric redshifts
        from R15 are both slightly higher, at 2.97 and 2.97, also reported at a
        high confidence level. The morphology of this source seems point-like,
        or at least very compact, in F336W. It is situated at the Northern tip
        of a more extended feature seen only in the non-ionizing filters.
        Another object is seen nearby to the NW (upper right) in all filters.
        This object has ID \#6770 in R15, where it has z\textsubscript{phot}
        \(\approx\) 1.55 with both their methods, but has no counterpart in I17.
        The two galaxies are well separated visually, but at a projected
        physical distance of \(\sim\) 5 kpc.\ at z=1.55, it is in principle
        possible that a source associated with this object could contaminate our
        source, but given the more obvious association with the source to the
        south with secure spectroscopic redshift, we believe it represents
        genuine LyC emission. The fluxes measured in the rest-frame LyC for this
        object are very high. It is quite young and with \(\log(M_*) \approx
        8.8\) in the lower mass end of our candidate list. It does however
        contain quite a lot of dust, which seems difficult to reconcile with
        such strong LyC emission. Indeed, while the \emph{absolute} escape
        fraction is 67\%, the \emph{relative} escape fraction is measured to
        876\%. While this is the highest of the derived relative escape
        fractions, as much as four galaxies have $f_{\text{esc}}^{\text{rel}} >
        100\%$. This is possible if the LyC emitting stars are seen through a
        dust column which is considerably lower than that of the stellar
        population dominating in the optical and IR filters and thus the stellar
        population models. It is well known that dust can be highly unevenly
        distributed in starburst galaxies \citep[see e.g.][Fig.~4 for an
        example]{menacho2021}. Additionally, it has been shown that the LyC
        output of a galaxy often is dominated by a few, perhaps as little as one
        or two, bright O or WR stars
        \citep{ramachandran2018a,ramachandran2018b,ramachandran2019,doran2013}.
        These stars can in some circumstances form in the outskirts of galaxies,
        e.g.\ as is the case of the Magellanic Bridge which contains at least 3
        bright O stars \citep{ramachandran2021}. The LyC detected in F336W-554
        does indeed originate in the outskirts of the galaxy, in a region that
        is quite faint in the optical and IR bands. 

\item[{\textbf{F336W-1013:}}] This galaxy has a spectroscopic redshift from I17
        of 2.98, which we have confirmed based on a Lyman \(\alpha\) line
        emission at \(\sim\) 4840 Å. The photometric redshift methods
        consistently converge on a redshift \(\sim 2\), which could be due to
        the assumption of a zero ionizing escape built-in to \texttt{BAGPIPES}.
        The galaxy was detected using F336W as detection filter. It has
        significant ionizing flux in F275W, and was also found in the source
        detection step, but fell for the color criterion and was rejected. The
        galaxy has an ionizing escape fraction of a moderate $15\pm10\%$ which,
        in combination with its redshift, leads to a strong IGM correction of
        the ionizing flux, making this galaxy the second strongest contributor
        to the ionizing volume emissivity $\epsilon_{900}$, delivering 10\% of
        the total amount of photons. The relative escape fraction of
        $f_{\text{esc}}^{\text{rel}} = 450\%$ is another example of a galaxy
        where the relative escape fraction indicates that the ionizing emission
        predominantly reaches us through lines of sight which are not typical
        for the general stellar population. Looking at the F275W cut-out of this
        galaxy in fig.~\ref{fig:stamps}, there is a brighter area to the W of
        the central object which, like the artifact next to F336W-1041, is not
        present in the non-PSF matched archival frames or the optical or IR
        frames. We performed the same tests as for F336W-1041 (see
        sect.~\ref{artifact}), measuring signal and noise in the non-PSF matched
        data, to check whether the main object, F336W-1013, could be a similar
        artifact, but found that it persisted at a very similar signal-to-noise
        as in the PSF matched data. The actual object is actually fainter than
        this possibly spurious, feature, meaning that at first glance it could
        easily be discarded as spurious, were it not for its complete spatial
        coincidence with the feature found in the non-ionizing bands. We thus
        believe it to be real, but the nature of the neighboring feature is
        still an open question. There is a hint of the same feature, with the
        same shape, in F336W, but not in any other band. However, none of these
        concerns are strong, and the Ly\(\alpha\) emission line has been
        identified with strong certainty, for which reason we have placed this
        galaxy among our most convincing candidates.

\item[{\textbf{F336W-1041:}}] This galaxy consistently converges on a
        photometric redshift of \(\sim\) 2.0--2.2 both in our computations and
        in R15, but has a bright Ly\(\alpha\) emission line at redshift z=3.33
        (see fig.~\ref{fig:OOIspecs}), which is also its designated redshift in
        I17, with a reported confidence level of 2; we speculate that as in the
        case of F336W-1013, the built-in assumption of zero ionizing escape in
        \texttt{BAGPIPES} might have skewed its solution towards a lower
        redshift. There are no strong emission lines consistent with the
        location of the one observed which correspond to redshifts between 2.0
        and 2.2. The high observed equivalent width of the emission feature
        limits which other lines it could conceivably be, as does the clearly
        asymmetrical line profile. 

        F336W-1041 has a relative escape fraction of 662\% and an absolute
        escape fraction of just over, but consistent with, 100\%. This means
        that not only must the line of sight to the LyC sources be completely
        free of internal stars and dust in the galaxy; also the LOS through the
        IGM must be essentially free of neutral Hydrogen. This is clearly a
        priori a very unlikely situation; however, as discussed above, our
        entire selection process is strongly favoring such a scenario among
        thousands of originally detected sources, and as argued above, the
        scenario is not implausible.
        
        The thumbnails in fig.~\ref{fig:stamps} show
        that the there is a nearby object to the NE of the galaxy. This
        companion has almost identical redshift in I17 (see
        Tab.~\ref{tab_redshifts}), which we interpret as the two being an
        interacting pair, and which means that any cross contamination between
        the two stems from effectively the same redshift and thus will not give
        rise to false detection in LyC. The flux is surprisingly strong in the
        LyC, leading to wonder whether we could be looking at a foreground
        interloper, but both UV, optical/IR imaging and Ly\(\alpha\) emission
        centroid coincide within a small distance, and the only other visible
        object in the vicinity has a spectroscopic redshift which would put its
        rest-frame LyC emission within the same filter as this object. The
        possibility of a near-perfect alignment with a faint foreground
        interloper exists, but is statistically very unlikely
        \citep{siana2010,nestor2013}. 

\end{description}

\subsection{Other candidates}
\label{sec:org5312824}
As discussed in sect. \ref{sec:org5cf08e5}, our tier-2 candidates are what we
consider likely, but not definite, candidates. Two of them have convincing and
consistent photometric redshifts which correspond to a rather strong LyC
emission, but lack spectroscopic redshift. One has a spectroscopic redshift, but
in one UV filter, it has SNR < 2, and the next straddles the Lyman break such
that the LyC detection there becomes model dependent. The last object has a
spectroscopic redshift of 2.98, at which LyC is detected by a solid margin, but
the three photometric redshifts are consistently \(\sim\) 2, which we find calls
for caution.

\begin{description}
\item[{\textbf{F275W-314:} }] This is the lowest-redshift and perhaps also the
  least convincing object in the selection. However, it does have a solid
  spectroscopic redshift, a >1\(\sigma\) detection in the rest-frame LyC filter
  F225W, and a flux in the Lyman edge straddling filter F275W which is difficult
  to reconcile with the inferred stellar population. However, this galaxy
  requires deeper imaging in F225W or a similar filter to be properly
  convincing.

\item[{\textbf{F275W-2055:}}] This galaxy is not included in I17, and we found
        no strong emission lines or convincing absorption lines in the MUSE
        spectrum. Our initial \texttt{BAGPIPES} run yielded \(z=2.07 \pm 0.08\),
        while the two methods of R15 yielded \(z=2.43\) and \(z=2.26\),
        respectively. However, in our initial SED modeling, we allowed
        metallicity to vary freely, which yielded an unbelievably high
        metallicity for this galaxy of \(Z\gtrsim 2 \times Z_{\odot}\). 

        In our second round of SED fitting, we constrained the metallicity to be
        equal to or below solar, which bumped up the inferred redshift to
        \(z=2.33 \pm 0.29\), better in line with the measurements of R15. A
        redshift of \(z=2.38\) is where <1\% of the integrated flux in the
        filter (for a flat spectrum in \(f_{\lambda}\) units) falls redward of
        the Lyman edge, but the observed flux in F336W for this galaxy becomes
        difficult to account for without invoking LyC escape already at
        redshifts \(z \gtrsim 2.2\), as can be seen in fig.~\ref{fig:sedfig}.
        Even at the original \(z=2.07\), the measured flux in F275W is
        inconsistent with the best-fit model if we assume no LyC escape within
        \(\sim\)2\(\sigma\). An even lower redshift, or a significantly bluer UV
        spectrum, could bring the measured SED within consistency with an
        absence of ionizing escape. We note again that the redshift is also
        downward biased in the fitting process by the built-in assumption of no
        ionizing escape. When the UV filters have lower statistical weight due
        to the larger aperture size and the metallicity is constrained to
        sub-solar values, the fit settles on a redshift entirely consistent with
        those from R15. In this redshift range, the segment based flux in this
        galaxy is inconsistent with an f\textsubscript{esc} of zero. The
        large-aperture F275W flux in this galaxy has a very large error bar, but
        as mentioned above, two effects inflate this artificially. One is the
        aperture size in itself, which includes more noise to the same signal;
        the other is that the noise is generally overestimated. We can get a
        rough estimate of the latter by looking at Fig.~\ref{fig:errors}, middle
        panel. Here, the green point cloud shows the ratio of RMS-based to
        MC-derived errors in the segments, plotted against the measured flux in
        the segments. The F275W-segment of F275W-2055 has a flux of roughly
        $0.8\times10^{20}$ ergs/s/cm$^{2}$/Å, meaning that the error on the flux
        was over estimated by a factor of \(\sim\) 3. We find this object
        convincing, but have assigned it Tier 2 due solely to the lack of a
        spectroscopic redshift.

\item[{\textbf{F336W-606:}}] The second of two candidates which has no
        spectroscopic redshift. For this galaxy, the R15 photometric redshifts
        are 2.50 and 2.69. Our original \texttt{BAGPIPES} redshift is 2.41 which
        however shifts to $z=2.73$ when the metallicity is constrained as
        described above. The galaxy has SNR > 2 in all three UV filters, but
        only picked up by our detection and filtering procedure when using F336W
        as detection filter. In F225W, it falls for the color criterion. It is
        unclear what is the reason why it is rejected when using F275W, but
        given the faintness of the source, it is conceivable that the structured
        noise could have been included in the image segment, and the difference
        in segment shape could dilute the SNR below the 2\(\sigma\) threshold.
        Morphologically, this is a complex object which consists of one major,
        bright, compact feature to the SW, and a collection of multiple fainter
        clumps to the NE (see Fig.~\ref{fig:stamps}). These features are all
        treated as one feature in R15, while we have based our SED modeling on a
        small segment of the NE feature collection, and even in the \(0\farcs8\)
        aperture, the two parts are separate. The combination of strong flux and
        consistent photometric redshifts make this an interesting candidate
        object, but the lack of a spectroscopic redshift is is a concern.
  
\end{description}

\subsection{Other LyC surveys in overlapping fields}
\label{sec:org0562031}
\subsubsection{Jones et al. 2021}
\label{sec:org6d58d4a}
\cite{jones2021} carry out a search for Lyman-continuum leakers in the GOODS
fields using the publicly available F275W observations from the HDUV survey
and archival spectroscopic catalogs to select galaxies in the redshift range
\(2.35 < z < 3.05\), in which the HST F275W filter probes only the
Lyman-continuum, while the adjacent F336W at least partially probes the
non-ionizing UV continuum. Given coordinates and redshifts from the HDUV
catalogs, they then lay down a \(2"\) circular aperture on the given locations in
F225W, and measure the flux in search of significant LyC detection, and then go
through various steps to control for contamination from interloping galaxies.
They find three candidate objects in GOODS-S, of which one, a galaxy at
\(z=2.678\) is contained in the UVUDF footprint. We found this object during
source detection in F275W, but it was subsequently discarded because the segment
had a signal-to-noise in F336W of only 1.7 (SNR criterion), and is 0.9 magnitude
fainter (color criterion). We have thus not reached the point of extracting MUSE
spectra or performing SED fitting for this object. We note that our method uses
small apertures, that only encompass the bright region in the detection image,
and relatively aggressively splits any clumpy source up in multiple segments.
This means that our method is quite robust to contamination from nearby objects,
while it is vulnerable towards configurations where the ionizing radiation may
emanate from regions that are faint in non-ionizing UV continuum.
\citeauthor{jones2021} use a larger aperture, encompassing the entire galaxy and
catching all flux within a radius of 1``, making this method more robust to
scenarios where there is an angular offset between ionizing and non-ionizing UV
radiation, but more vulnerable towards inaccuracies in background subtraction,
contamination from adjacent objects, etc. \citeauthor{jones2021} do not find our
object F336W-554; It is unclear why, as these authors have used the catalogs of
I17 for preselection of candidates, and the object is found within the redshift
ranges that they probe.\\

\subsubsection{Saxena et al. 2021} \label{sec:org669876e} \cite{saxena2021}
search for LyC leakers in the  CDF-S, a superset of the HUDF. In their  work,
they look  at  galaxies  at \(3.1  <  z  < 3.4\)  pre-selected  from photometry
or  MUSE-DEEP, find spectroscopic  redshifts in the  VANDELS catalogs
\citep{mclure2018}, and look  for LyC emission in archival  photometric data.
They find 11 LyC leaker candidates in the  greater field, of which 4 are
contained in the  HUDF,  including  the  one  we label  F336W-189  and  these
authors  label CDFS-12448.  Interestingly,  this  object  is   the  only
overlap  between  our candidates and those  of \citeauthor{saxena2021}: Our
method did  not yield either of  their  other 3  objects  inside  the  footprint
of the  UVUDF:  CDFS-13385, CDFS-15718, or CDFS-16444.

\begin{description}
\item[{\textbf{CDFS 12448}: }] As discussed above, this is the same object that
  we denote F336W--189, and we agree that it is a strong Lyman continuum leaker
  candidate.

\item[{\textbf{CDFS 13385:}}] We detected this galaxy in all the three detection
  passes. In F225W and F275W, it fell for the color criterion discussed in
  sect.~\ref{sec:uvcolors}, meaning that the detection filter was brighter in AB
  magnitudes than the next redward filter. When detected using F336W, it passed
  these criteria and was selected for SED fitting. Our SED fitting yielded a
  photometric redshift of \(z = 0.46^{+0.009}_{-0.26}\). We compared this to
  \cite{rafelski15_uvudf} who reported the redshifts \(z = 0.09\) and
  \(z = 0.20\) with their two independent methods. In \cite{inami2017}, the
  object's redshift is reported with a confidence level of 2 out of 3, or
  ``fairly strong'', based on identification of Lyman-\(\alpha\) in emission.
  However, given the SED and the agreement between the three photometric
  redshifts, we find it likely that the redshift in \cite{inami2017} is based on
  a misclassified [\ion{O}{ii}] \(\lambda \lambda\) 3727,3729 doublet which at a
  redshift of \(\sim\)0.44 would fall at the wavelength of Lyman \(\alpha\) at
  the redshift reported by \citeauthor{inami2017}. At this redshift, the doublet
  would have a peak separation of \(\sim\) 4 Å, corresponding to a peak
  separation of \(\sim\) 225 km/s, which could reasonably be mistaken for a
  \(z=3.4\) Ly\(\alpha\) profile by an automated survey.

\item[{\textbf{CDFS 15718}:}] This galaxy is not detected in the HST UV filters
  in the UVUDF, despite these observations being quite deep. At its redshift of
  \(z\approx3.44\), the Lyman edge falls around 4050 \AA{}, close to the red
  edge of the NB396 narrow band filter at the 2.2M telescope at La Silla, in
  which they have observed the reported LyC emission. F336W, on the other hand,
  falls blueward of NB396, between \(\sim\) 3100 and \(\sim\) 3700 \AA{}. It is
  possible that a Lyman Limit system exists between the galaxy's redshift of
  3.44 and z\(\sim\)3.05, where LyC would enter the red side of F336W. This
  could conceivably allow a window of LyC to reach Earth and be be detected in
  NB396 but not F336W.

\item[{\textbf{CDFS 16444}: }] We have not detected this galaxy in F336W either.
  As with CDFS 15718, \citeauthor{saxena2021} report this galaxy detected in
  Lyman-continuum in the NB396 filter. This galaxy is located at \(z=3.13\),
  leaving the window of \(3.05<z<3.14\) for a possible Lyman Limit system to
  absorb escaping LyC on its blue side.
\end{description}

Conversely,  the method  of \citeauthor{saxena2021}  did not  yield either  of the
objects  labeled F336W-1041  or F336W-554  in this  work. The  latter is  easily
explained,  as its  spectroscopic redshift  is outside  their redshift  range of
interest. F336W-1041, on  the other hand, is within the  redshift range, but not
listed among their candidates.

\subsubsection{Redshift distribution}
\label{sec:org86f4514}
Of the three redshift bins defined by the wavelengths of the detection filters,
our leaker candidates were found predominantly in the highest redshift bin, with
no candidates found in the lowest redshift bin. The latter can be at least
partially explained by the large overlap in transmission of F225W and F275W; the
volume unique to F225W is small, and F275W is deeper. If the Lyman edge of a
galaxy falls within F225W, the distinction between ionizing and non-ionizing
radiation becomes difficult and model dependent. If on the other hand it falls
entirely redward of this filter, only a narrow redshift window exists before it
also drops out of F275W.

It is somewhat surprising that we find the leaking candidates so predominantly
at higher redshift, given that the IGM becomes more opaque at the higher
redshifts. The co-moving volumes probed by the two highest-redshift filters are
the same order of magnitude, but the higher neutral IGM fraction in the highest
redshift bin (\(3.05 < z \lesssim 4\)) should, if average galaxy properties were
unchanged, result in a substantially larger number of LyC emitting galaxies in
the lower redshift bin.
Galaxy properties are however not unchanged over redshifts. While the cosmic
star formation rate is largely unchanged over this redshift range
\citep{madau2014}, the LAE fraction does rise \citep{stark2010,hayes2011}; which
is generally believed to be a good statistical tracer of Lyman Continuum
emission \citep[e.g.][and references therein]{matthee2022,naidu2022}. Thus,
the numbers of intrinsic leakers, and their average escape fractions, is
expected to rise with redshift.

In addition, there is an observational bias owing to F336W being considerably
deeper than the other two FUV filters, and the ACS F435W filter used to retain
detections is significantly deeper yet.

\subsection{Discrepancies in dust attenuation of ionizing and nonionizing filters}
\label{sec:fescdisc}

We have found in in sect.~\ref{sec:fescape} that in order for our derived escape
fractions to be physical, we must assume that the stars dominating the LyC
output are seen through a substantially thinner dust cover than the larger
population emitting the bulk of restframe 1500 Å photons. After IGM correction,
we report relative escape fractions of several hundred percent in some of our
candidates. However, all absolute escape fractions found are consistent with
physical values, so the relative escape fractions are only unphysical if
requiring that the LyC sources and the general stellar population be seen
through the same dust cover; the situation is different if we are seeing the LyC
emitting stars through a more modest dust cover. As touched upon in that
section, this is not an unreasonable assumption, given that dust has been found
to be highly clumpy in nearby galaxies such as Haro 11 \citep[][Rivera-Thorsen
et al. in prep.]{ostlin2015,ostlin2021,menacho2019}, and given that the LyC
output of star-forming galaxies can be dominated by a small number of very
luminous O stars, sometimes as few as one or two. For these, stochastic
variations in the line-of-sight HI and dust column can be expected to play a
substantially larger role than for the larger surrounding, LyC-faint population.
This uncertainty in the relative dust attenuations at $\lambda < 900 Å
\text{vs.} \lambda \approx 1500 Å$ should be kept in mind when drawing
conclusions about cosmic LyC output based on measured relative escape fractions.

\subsection{F336W artifacts near F336W-1013 and F336W-1041           \label{artifact}}
\label{sec:org2478580}
As is seen in the thumbnails of fig.~\ref{fig:stamps}, a feature is visible in
F336W to the N of the central source which is somewhat removed from the source
seen to the NE of the central source in the optical and IR filters. The features
are shown in a larger size in Fig.~\ref{fig:mergerpair}. This feature was picked
up and assigned the ID F336W-1043 by our source detection process, but rejected
in subsequent steps due to the lack of a source in the same location in the
optical filters. It did however raise concerns as to whether faint foreground
interloper was present in this location. However, when returning to the non-PSF
matched archival data, the source was not present, and neither is it present in
these data when applying a smoothing kernel. We speculate that this might be an
artifact of the PSF matching process, but have no explanation for it.
To test whether a similar effect might have led to the feature detected as
F336W-1041, we measured flux and noise in an aperture at this location in the
non-PSF matched image, but found that both flux and error estimate were similar
to what we found in the PSF matched image.

A similar artifact is present in the thumbnail of F336W-1013, which was likewise
not found in the non-PSF-matched images, while the flux measured in the source
itself was consistent between the raw and PSF matched images.

\section{Summary and conclusions                          \label{sec:summary}}
\label{sec:orgee5ee3c}
In this  work, we have  investigated how much  more ionizing radiation,  if any,
would be picked  up in a study completely eliminating  preselection, compared to
previous studies who have all relied  on a range of photometric or spectroscopic
preselection criteria for candidate identification.

To  this end,  we have  performed a  ``bottom-up'' search,  in which  we began
by identifying all sources in  the Hubble Ultra Deep Field in  the UV filters
which correspond to rest-frame  LyC emission in the  redshift range \(2 < z  <
3.5\). We then  performed photometric  measurements in  small apertures  defined
by  these UV-detected features in PSF-matched imaging data using all 11
\emph{HST} filters in the HUDF,  and use  these  to apply  color-  and SNR
based  selection criteria  to eliminate  sources which  were not  consistent
with  being LyC  emitters in  the appropriate redshift ranges.

We have performed SED fitting on a  number of candidates which were not rejected
by the  initial filtering  criteria, found  inferred photometric  redshifts, and
have compared these to catalogs  of photometric and spectroscopic redshifts from
the  literature  \citep{rafelski15_uvudf,inami2017}  to  find  candidate  objects,
combined  this   with  visual   inspections  looking  for   possible  foreground
interlopers to finally select a sample of seven galaxies divided in a tier-1 and
a  tier-2 sample,  based on  how convincingly  they met  the criteria  of having
secure redshift  measurements and convincingly  detected flux in  the rest-frame
LyC.

For this sample of galaxies, we have  re-extracted fluxes in all available of the
11  HUDF filters,  this  time  using fixed  circular  apertures of  \(0\farcs8\),
corresponding to the fiducial seeing adopted  in I17, which besides being useful
for  direct comparison  with ground-based  studies also  strikes a  good balance
between covering the majority of a  typical galaxy in this redshift range, while
being small  enough to still make  it unlikely to catch  contaminating flux from
nearby  objects. From  these  fluxes, we  have performed  a  more stringent  SED
modeling,  from  which  we  derived  the reported  physical  properties  of  the
galaxies. From the SED modeled stellar populations, we have derived intrinsic
ionizing output and absolute and relative escape fractions for all candidates.

Based on these measurements and models, we have computed the contribution to the
metagalactic ionizing background  from star formation in our  survey volume, and
found that cosmic variance  for this volume and in the relevant  mass range is a
modest \(\lesssim 20\%\) in flux. We have  compared the results of our methods to
two   other   recent  searches   for   LyC   leakers  in   overlapping   fields,
\citep{jones2021,saxena2021}, and discuss which  strengths and vulnerabilities the
various methods may have.

\begin{enumerate} 

\item From the initial total 6512 detections (of which an unknown number
        correspond to the same object in multiple filters) in the three
        detection filters, covering the redshift range \(2 < z < 3.5\), we have
        found a selection of 7 candidate objects. None of these candidates were
        originally detected in F225W.

\item Compared to other recent studies searching for LyC leakers in overlapping
        survey volumes \citep{jones2021,saxena2021} using various preselection
        methods, our work has found candidates not identified by the other
        authors, but also either missed or rejected candidates identified by
        these authors. The candidate found by \citeauthor{jones2021} was
        rejected because the S/N in the segment used for photometry in this work
        was too low, even if it was large enough in the larger aperture. This
        highlights that our method relies on the ionizing and nonionizing UV
        continuum being coincident and compact. This is a choice informed by the
        fact that the hot stars emitting strongest in the ionizing continuum
        also are bright in the non-ionizing UV, and that LyC emission in
        galaxies often is dominated by a few, sometimes just one or two, hot O
        stars
        \citep{doran2013,ramachandran2018a,ramachandran2018b,ramachandran2019}.
        However, O stars have also been observed in more tenuous environment
        like the Magellanic Bridge \citep{ramachandran2021}, where the chance of
        escape may be larger than in brighter central objects. Two candidates
        reported by \citet{saxena2021} were not detected in the rest-frame LyC
        in F336W, which is somewhat bluer than the LyC filter of these authors.
        We speculate that this might be explained by the fact that these authors
        use a filter for LyC which is narrower and slightly redder than F336W,
        placing it closer to the Lyman edge, where it could conceivably detect a
        rest-frame \(\sim\) 850 Å signal, while the rest-frame \(\sim\)
        750--800Å signal which would be observed in F336W could get absorbed by
        the IGM.

\item We find very high relative escape fractions in all candidate galaxies, up
        to $\sim 450\%$ after IGM correction, and likely substantially more if
        correcting by a more realistic dust model. Such values are tempting to
        dismiss as unphysical. However, all found leaker candidates have
        absolute escape fractions consistent with unity; the high relative
        escape fractions are due to correction for the dust cover inferred from
        the SED fits of the general stellar population, based on the assumption
        that the LyC leaking stars are seen through a dust layer identical to
        that covering the general stellar population. Dust is often distributed
        highly unevenly in starburst galaxies, and their LyC output may be
        dominated by a small number of very bright stars, suggesting that these
        can indeed be seen through channels of lower dust content, perhaps
        practically zero. This calls for caution when interpreting the relative
        escape fractions often reported in the literature. 

\item We find that for some of our candidate objects, the stars contributing the
        bulk of LyC photons must be subject to substantially lower dust
        attenuation than the stellar population dominating at rest-frame 1500 Å.
        We have argued that this could be a believable scenario based on the
        strong patchiness of dust found in studies of local-Universe starburst
        galaxies; on previous findings that a very small number of stars can
        dominate the ionizing output of a galaxy, and that our selection
        criteria have favored just such a scenario strongly. We argue that this
        can have consequences for estimates of cosmic ionizing emissivity based
        on estimates of relative escape fractions.

\item We have computed a value of the cosmic ionizing emissivity of
        log\textsubscript{10}(\(\epsilon\)\textsubscript{\(\nu\)})\(=
        25.32^{+0.25}_{-0.21}\) assuming all our candidates are bona fide LyC
        leakers, and \(25.29^{+0.27}_{-0.22}\) assuming only the tier 1
        candidates are real leakers. We have based these numbers on the
        assumption that IGM lines of sight which lead to
        f\textsubscript{esc,abs} \(\le\) 100\% are valid basis for Monte Carlo
        sampling of the IGM.

\item The derived contribution
        log\textsubscript{10}(\(\epsilon\)\textsubscript{\(\nu\)}) to the
        metagalactic ionizing background made by these galaxies is consistent
        with the values found from indirect measurements of the ionization state
        of the IGM done by \citet{becker.bolton2013} without invoking bright
        quasars. It is also consistent with the values found for the redshift
        range \(2.35 < z < 3.05\) found by \citet{jones2021}. Our findings thus
        support the claim that star formation can indeed produce the amount of
        ionizing photons necessary to account for the ionization degree of the
        IGM during the epoch of \emph{Cosmic Noon}.

\end{enumerate}

Comparison with studies using widely adopted candidate preselection methods
shows that the method in this work complements the methods using preselection:
It can identify candidates that are otherwise overlooked by these methods, but
can also miss candidates found by other methods. We therefore recommend the
community that searches based on the general method of this work be carried out
in the other large \emph{Hubble} legacy fields such as the wider CANDELS fields.

\begin{acknowledgements} The authors thank the anonymous reviewer for kind and
constructive criticism which led to a substantial improvement to the quality of
this paper. MH is a fellow of the Knut \& Alice Wallenberg Foundation.
\end{acknowledgements} 

\bibliography{AllPapers}

\begin{thebibliography}{108}
\expandafter\ifx\csname natexlab\endcsname\relax\def\natexlab#1{#1}\fi

\bibitem[{{Alavi} {et~al.}(2020){Alavi}, {Colbert}, {Teplitz}, {Siana},
  {Scarlata}, {Rutkowski}, {Mehta}, {Henry}, {Dai}, {Haardt}, \&
  {Bagley}}]{alavi2020}
{Alavi}, A., {Colbert}, J., {Teplitz}, H.~I., {et~al.} 2020, \apj, 904, 59

\bibitem[{{Bacon} {et~al.}(2015){Bacon}, {Brinchmann}, {Richard}, {Contini},
  {Drake}, {Franx}, {Tacchella}, {Vernet}, {Wisotzki}, {Blaizot}, {Bouch{\'e}},
  {Bouwens}, {Cantalupo}, {Carollo}, {Carton}, {Caruana}, {Cl{\'e}ment},
  {Dreizler}, {Epinat}, {Guiderdoni}, {Herenz}, {Husser}, {Kamann}, {Kerutt},
  {Kollatschny}, {Krajnovic}, {Lilly}, {Martinsson}, {Michel-Dansac},
  {Patricio}, {Schaye}, {Shirazi}, {Soto}, {Soucail}, {Steinmetz}, {Urrutia},
  {Weilbacher}, \& {de Zeeuw}}]{bacon2015}
{Bacon}, R., {Brinchmann}, J., {Richard}, J., {et~al.} 2015, \aap, 575, A75

\bibitem[{{Bassett} {et~al.}(2019){Bassett}, {Ryan-Weber}, {Cooke}, {Diaz},
  {Nanayakkara}, {Yuan}, {Spitler}, {Me{\v{s}}tri{\'c}}, {Garel}, {Sawicki},
  {Gwyn}, \& {Golob}}]{bassett2019}
{Bassett}, R., {Ryan-Weber}, E.~V., {Cooke}, J., {et~al.} 2019, \mnras, 483,
  5223

\bibitem[{{Bassett} {et~al.}(2021){Bassett}, {Ryan-Weber}, {Cooke},
  {Me{\v{s}}tri{\'c}}, {Kakiichi}, {Prichard}, \& {Rafelski}}]{bassett2021}
{Bassett}, R., {Ryan-Weber}, E.~V., {Cooke}, J., {et~al.} 2021, \mnras, 502,
  108

\bibitem[{{Bassett} {et~al.}(2022){Bassett}, {Ryan-Weber}, {Cooke},
  {Me{\v{s}}tri{\'c}}, {Prichard}, {Rafelski}, {Iwata}, {Sawicki}, {Gwyn}, \&
  {Arnouts}}]{bassett2022}
{Bassett}, R., {Ryan-Weber}, E.~V., {Cooke}, J., {et~al.} 2022, \mnras, 511,
  5730

\bibitem[{{Becker} \& {Bolton}(2013)}]{becker.bolton2013}
{Becker}, G.~D. \& {Bolton}, J.~S. 2013, \mnras, 436, 1023

\bibitem[{{Becker} {et~al.}(2021){Becker}, {D'Aloisio}, {Christenson}, {Zhu},
  {Worseck}, \& {Bolton}}]{becker2021}
{Becker}, G.~D., {D'Aloisio}, A., {Christenson}, H.~M., {et~al.} 2021, \mnras,
  508, 1853

\bibitem[{{Beckwith} {et~al.}(2006){Beckwith}, {Stiavelli}, {Koekemoer},
  {Caldwell}, {Ferguson}, {Hook}, {Lucas}, {Bergeron}, {Corbin}, {Jogee},
  {Panagia}, {Robberto}, {Royle}, {Somerville}, \& {Sosey}}]{beckwith2006}
{Beckwith}, S. V.~W., {Stiavelli}, M., {Koekemoer}, A.~M., {et~al.} 2006, \aj,
  132, 1729

\bibitem[{{Behrens} {et~al.}(2014){Behrens}, {Dijkstra}, \&
  {Niemeyer}}]{behrens2014}
{Behrens}, C., {Dijkstra}, M., \& {Niemeyer}, J.~C. 2014, \aap, 563, A77

\bibitem[{{Ben{\'\i}tez}(2000)}]{benitez2000}
{Ben{\'\i}tez}, N. 2000, \apj, 536, 571

\bibitem[{{Ben{\'\i}tez} {et~al.}(2004){Ben{\'\i}tez}, {Ford}, {Bouwens},
  {Menanteau}, {Blakeslee}, {Gronwall}, {Illingworth}, {Meurer}, {Broadhurst},
  {Clampin}, {Franx}, {Hartig}, {Magee}, {Sirianni}, {Ardila}, {Bartko},
  {Brown}, {Burrows}, {Cheng}, {Cross}, {Feldman}, {Golimowski}, {Infante},
  {Kimble}, {Krist}, {Lesser}, {Levay}, {Martel}, {Miley}, {Postman}, {Rosati},
  {Sparks}, {Tran}, {Tsvetanov}, {White}, \& {Zheng}}]{benitez2004}
{Ben{\'\i}tez}, N., {Ford}, H., {Bouwens}, R., {et~al.} 2004, \apjs, 150, 1

\bibitem[{{Bergvall} {et~al.}(2006){Bergvall}, {Zackrisson}, {Andersson},
  {Arnberg}, {Masegosa}, \& {{\"O}stlin}}]{bergvall2006}
{Bergvall}, N., {Zackrisson}, E., {Andersson}, B.-G., {et~al.} 2006, \aap, 448,
  513

\bibitem[{{Bian} {et~al.}(2017){Bian}, {Fan}, {McGreer}, {Cai}, \&
  {Jiang}}]{bian2017}
{Bian}, F., {Fan}, X., {McGreer}, I., {Cai}, Z., \& {Jiang}, L. 2017, \apjl,
  837, L12

\bibitem[{{Borthakur} {et~al.}(2014){Borthakur}, {Heckman}, {Leitherer}, \&
  {Overzier}}]{borthakur2014}
{Borthakur}, S., {Heckman}, T.~M., {Leitherer}, C., \& {Overzier}, R.~A. 2014,
  Science, 346, 216

\bibitem[{{Bosman} {et~al.}(2022){Bosman}, {Davies}, {Becker}, {Keating},
  {Davies}, {Zhu}, {Eilers}, {D'Odorico}, {Bian}, {Bischetti}, {Cristiani},
  {Fan}, {Farina}, {Haehnelt}, {Hennawi}, {Kulkarni}, {Mesinger}, {Meyer},
  {Onoue}, {Pallottini}, {Qin}, {Ryan-Weber}, {Schindler}, {Walter}, {Wang}, \&
  {Yang}}]{bosman2022}
{Bosman}, S. E.~I., {Davies}, F.~B., {Becker}, G.~D., {et~al.} 2022, \mnras,
  514, 55

\bibitem[{{Bouwens} {et~al.}(2011){Bouwens}, {Illingworth}, {Oesch},
  {Labb{\'e}}, {Trenti}, {van Dokkum}, {Franx}, {Stiavelli}, {Carollo},
  {Magee}, \& {Gonzalez}}]{bouwens2011}
{Bouwens}, R.~J., {Illingworth}, G.~D., {Oesch}, P.~A., {et~al.} 2011, \apj,
  737, 90

\bibitem[{Bradley {et~al.}(2020)Bradley, Sip{\H o}cz, Robitaille, Tollerud,
  Vin{\'{\i}}cius, Deil, Barbary, Wilson, Busko, G{\"u}nther, Cara, Conseil,
  Bostroem, Droettboom, Bray, Bratholm, Lim, Barentsen, Craig, Pascual, Perren,
  Greco, Donath, de~Val-Borro, Kerzendorf, Bach, Weaver, D'Eugenio, Souchereau,
  \& Ferreira}]{photutils2020}
Bradley, L., Sip{\H o}cz, B., Robitaille, T., {et~al.} 2020, astropy/photutils:
  1.0.0

\bibitem[{{Brammer} {et~al.}(2008){Brammer}, {van Dokkum}, \&
  {Coppi}}]{brammer2008eazy}
{Brammer}, G.~B., {van Dokkum}, P.~G., \& {Coppi}, P. 2008, \apj, 686, 1503

\bibitem[{{Bruzual} \& {Charlot}(2003)}]{bruzual2003}
{Bruzual}, G. \& {Charlot}, S. 2003, \mnras, 344, 1000

\bibitem[{{Buchner} {et~al.}(2014){Buchner}, {Georgakakis}, {Nandra}, {Hsu},
  {Rangel}, {Brightman}, {Merloni}, {Salvato}, {Donley}, \&
  {Kocevski}}]{buchner2014}
{Buchner}, J., {Georgakakis}, A., {Nandra}, K., {et~al.} 2014, \aap, 564, A125

\bibitem[{{Calzetti} {et~al.}(2000){Calzetti}, {Armus}, {Bohlin}, {Kinney},
  {Koornneef}, \& {Storchi-Bergmann}}]{calzetti2000}
{Calzetti}, D., {Armus}, L., {Bohlin}, R.~C., {et~al.} 2000, \apj, 533, 682

\bibitem[{{Carnall} {et~al.}(2018){Carnall}, {McLure}, {Dunlop}, \&
  {Dav{\'e}}}]{carnall2018}
{Carnall}, A.~C., {McLure}, R.~J., {Dunlop}, J.~S., \& {Dav{\'e}}, R. 2018,
  \mnras, 480, 4379

\bibitem[{{Chisholm} {et~al.}(2018){Chisholm}, {Gazagnes}, {Schaerer},
  {Verhamme}, {Rigby}, {Bayliss}, {Sharon}, {Gladders}, \&
  {Dahle}}]{chisholm2018a}
{Chisholm}, J., {Gazagnes}, S., {Schaerer}, D., {et~al.} 2018, \aap, 616, A30

\bibitem[{{Coe} {et~al.}(2006){Coe}, {Ben{\'\i}tez}, {S{\'a}nchez}, {Jee},
  {Bouwens}, \& {Ford}}]{coe2006}
{Coe}, D., {Ben{\'\i}tez}, N., {S{\'a}nchez}, S.~F., {et~al.} 2006, \aj, 132,
  926

\bibitem[{{Cooke} {et~al.}(2014){Cooke}, {Ryan-Weber}, {Garel}, \&
  {D{\'\i}az}}]{cooke2014}
{Cooke}, J., {Ryan-Weber}, E.~V., {Garel}, T., \& {D{\'\i}az}, C.~G. 2014,
  \mnras, 441, 837

\bibitem[{{Cowie} {et~al.}(2009){Cowie}, {Barger}, \& {Trouille}}]{cowie2009}
{Cowie}, L.~L., {Barger}, A.~J., \& {Trouille}, L. 2009, \apj, 692, 1476

\bibitem[{{Davies} {et~al.}(2021){Davies}, {Bosman}, {Furlanetto}, {Becker}, \&
  {D'Aloisio}}]{davies2021}
{Davies}, F.~B., {Bosman}, S. E.~I., {Furlanetto}, S.~R., {Becker}, G.~D., \&
  {D'Aloisio}, A. 2021, \apjl, 918, L35

\bibitem[{{Dayal} {et~al.}(2020){Dayal}, {Volonteri}, {Choudhury}, {Schneider},
  {Trebitsch}, {Gnedin}, {Atek}, {Hirschmann}, \& {Reines}}]{dayal2020}
{Dayal}, P., {Volonteri}, M., {Choudhury}, T.~R., {et~al.} 2020, \mnras, 495,
  3065

\bibitem[{{de Barros} {et~al.}(2016){de Barros}, {Vanzella}, {Amor{\'{\i}}n},
  {Castellano}, {Siana}, {Grazian}, {Suh}, {Balestra}, {Vignali}, {Verhamme},
  {Zamorani}, {Mignoli}, {Hasinger}, {Comastri}, {Pentericci},
  {P{\'e}rez-Montero}, {Fontana}, {Giavalisco}, \& {Gilli}}]{debarros2016}
{de Barros}, S., {Vanzella}, E., {Amor{\'{\i}}n}, R., {et~al.} 2016, \aap, 585,
  A51

\bibitem[{{Doran} {et~al.}(2013){Doran}, {Crowther}, {de Koter}, {Evans},
  {McEvoy}, {Walborn}, {Bastian}, {Bestenlehner}, {Gr{\"a}fener}, {Herrero},
  {K{\"o}hler}, {Ma{\'\i}z Apell{\'a}niz}, {Najarro}, {Puls}, {Sana},
  {Schneider}, {Taylor}, {van Loon}, \& {Vink}}]{doran2013}
{Doran}, E.~I., {Crowther}, P.~A., {de Koter}, A., {et~al.} 2013, \aap, 558,
  A134

\bibitem[{{Ellis} {et~al.}(2013){Ellis}, {McLure}, {Dunlop}, {Robertson},
  {Ono}, {Schenker}, {Koekemoer}, {Bowler}, {Ouchi}, {Rogers}, {Curtis-Lake},
  {Schneider}, {Charlot}, {Stark}, {Furlanetto}, \& {Cirasuolo}}]{ellis2013}
{Ellis}, R.~S., {McLure}, R.~J., {Dunlop}, J.~S., {et~al.} 2013, \apjl, 763, L7

\bibitem[{{Fan} {et~al.}(2006){Fan}, {Carilli}, \& {Keating}}]{fan2006}
{Fan}, X., {Carilli}, C.~L., \& {Keating}, B. 2006, Annual Review of Astronomy
  and Astrophysics, 44, 415

\bibitem[{{Faucher-Gigu{\`e}re}(2020)}]{fauchergiguere2020}
{Faucher-Gigu{\`e}re}, C.-A. 2020, \mnras, 493, 1614

\bibitem[{{Ferland} {et~al.}(2017){Ferland}, {Chatzikos}, {Guzm{\'a}n},
  {Lykins}, {van Hoof}, {Williams}, {Abel}, {Badnell}, {Keenan}, {Porter}, \&
  {Stancil}}]{ferland2017}
{Ferland}, G.~J., {Chatzikos}, M., {Guzm{\'a}n}, F., {et~al.} 2017, \rmxaa, 53,
  385

\bibitem[{{Feroz} {et~al.}(2019){Feroz}, {Hobson}, {Cameron}, \&
  {Pettitt}}]{feroz2019}
{Feroz}, F., {Hobson}, M.~P., {Cameron}, E., \& {Pettitt}, A.~N. 2019, The Open
  Journal of Astrophysics, 2, 10

\bibitem[{{Finkelstein} {et~al.}(2019){Finkelstein}, {D'Aloisio},
  {Paardekooper}, {Ryan}, {Behroozi}, {Finlator}, {Livermore}, {Upton
  Sanderbeck}, {Dalla Vecchia}, \& {Khochfar}}]{finkelstein2019}
{Finkelstein}, S.~L., {D'Aloisio}, A., {Paardekooper}, J.-P., {et~al.} 2019,
  \apj, 879, 36

\bibitem[{{Fletcher} {et~al.}(2019){Fletcher}, {Tang}, {Robertson}, {Nakajima},
  {Ellis}, {Stark}, \& {Inoue}}]{fletcher2018}
{Fletcher}, T.~J., {Tang}, M., {Robertson}, B.~E., {et~al.} 2019, \apj, 878, 87

\bibitem[{{Flury} {et~al.}(2022){Flury}, {Jaskot}, {Ferguson}, {Worseck},
  {Makan}, {Chisholm}, {Saldana-Lopez}, {Schaerer}, {McCandliss}, {Wang},
  {Ford}, {Heckman}, {Ji}, {Giavalisco}, {Amorin}, {Atek}, {Blaizot},
  {Borthakur}, {Carr}, {Castellano}, {Cristiani}, {De Barros}, {Dickinson},
  {Finkelstein}, {Fleming}, {Fontanot}, {Garel}, {Grazian}, {Hayes}, {Henry},
  {Mauerhofer}, {Micheva}, {Oey}, {Ostlin}, {Papovich}, {Pentericci},
  {Ravindranath}, {Rosdahl}, {Rutkowski}, {Santini}, {Scarlata}, {Teplitz},
  {Thuan}, {Trebitsch}, {Vanzella}, {Verhamme}, \& {Xu}}]{flury2022a}
{Flury}, S.~R., {Jaskot}, A.~E., {Ferguson}, H.~C., {et~al.} 2022, \apjs, 260,
  1

\bibitem[{{Grazian} {et~al.}(2016){Grazian}, {Giallongo}, {Gerbasi}, {Fiore},
  {Fontana}, {Le F{\`e}vre}, {Pentericci}, {Vanzella}, {Zamorani}, {Cassata},
  {Garilli}, {Le Brun}, {Maccagni}, {Tasca}, {Thomas}, {Zucca},
  {Amor{\'{\i}}n}, {Bardelli}, {Cassar{\`a}}, {Castellano}, {Cimatti},
  {Cucciati}, {Durkalec}, {Giavalisco}, {Hathi}, {Ilbert}, {Lemaux}, {Paltani},
  {Ribeiro}, {Schaerer}, {Scodeggio}, {Sommariva}, {Talia}, {Tresse},
  {Vergani}, {Bonchi}, {Boutsia}, {Capak}, {Charlot}, {Contini}, {de la Torre},
  {Dunlop}, {Fotopoulou}, {Guaita}, {Koekemoer}, {L{\'o}pez-Sanjuan},
  {Mellier}, {Merlin}, {Paris}, {Pforr}, {Pilo}, {Santini}, {Scoville},
  {Taniguchi}, \& {Wang}}]{grazian2016}
{Grazian}, A., {Giallongo}, E., {Gerbasi}, R., {et~al.} 2016, \aap, 585, A48

\bibitem[{{Grazian} {et~al.}(2017){Grazian}, {Giallongo}, {Paris}, {Boutsia},
  {Dickinson}, {Santini}, {Windhorst}, {Jansen}, {Cohen}, {Ashcraft},
  {Scarlata}, {Rutkowski}, {Vanzella}, {Cusano}, {Cristiani}, {Giavalisco},
  {Ferguson}, {Koekemoer}, {Grogin}, {Castellano}, {Fiore}, {Fontana},
  {Marchi}, {Pedichini}, {Pentericci}, {Amor{\'\i}n}, {Barro}, {Bonchi},
  {Bongiorno}, {Faber}, {Fumana}, {Galametz}, {Guaita}, {Kocevski}, {Merlin},
  {Nonino}, {O'Connell}, {Pilo}, {Ryan}, {Sani}, {Speziali}, {Testa}, {Weiner},
  \& {Yan}}]{grazian2017}
{Grazian}, A., {Giallongo}, E., {Paris}, D., {et~al.} 2017, \aap, 602, A18

\bibitem[{{Haardt} \& {Madau}(2012)}]{haardt2012}
{Haardt}, F. \& {Madau}, P. 2012, \apj, 746, 125

\bibitem[{{Hayes} {et~al.}(2011){Hayes}, {Schaerer}, {{\"O}stlin}, {Mas-Hesse},
  {Atek}, \& {Kunth}}]{hayes2011}
{Hayes}, M., {Schaerer}, D., {{\"O}stlin}, G., {et~al.} 2011, \apj, 730, 8

\bibitem[{{Inami} {et~al.}(2017){Inami}, {Bacon}, {Brinchmann}, {Richard},
  {Contini}, {Conseil}, {Hamer}, {Akhlaghi}, {Bouch{\'e}}, {Cl{\'e}ment},
  {Desprez}, {Drake}, {Hashimoto}, {Leclercq}, {Maseda}, {Michel-Dansac},
  {Paalvast}, {Tresse}, {Ventou}, {Kollatschny}, {Boogaard}, {Finley},
  {Marino}, {Schaye}, \& {Wisotzki}}]{inami2017}
{Inami}, H., {Bacon}, R., {Brinchmann}, J., {et~al.} 2017, \aap, 608, A2

\bibitem[{{Inoue} {et~al.}(2014){Inoue}, {Shimizu}, {Iwata}, \&
  {Tanaka}}]{inoue2014}
{Inoue}, A.~K., {Shimizu}, I., {Iwata}, I., \& {Tanaka}, M. 2014, \mnras, 442,
  1805

\bibitem[{{Izotov} {et~al.}(2016{\natexlab{a}}){Izotov}, {Orlitov{\'a}},
  {Schaerer}, {Thuan}, {Verhamme}, {Guseva}, \& {Worseck}}]{izotov2016nature}
{Izotov}, Y.~I., {Orlitov{\'a}}, I., {Schaerer}, D., {et~al.}
  2016{\natexlab{a}}, \nat, 529, 178

\bibitem[{{Izotov} {et~al.}(2016{\natexlab{b}}){Izotov}, {Schaerer}, {Thuan},
  {Worseck}, {Guseva}, {Orlitov{\'a}}, \& {Verhamme}}]{izotov2016}
{Izotov}, Y.~I., {Schaerer}, D., {Thuan}, T.~X., {et~al.} 2016{\natexlab{b}},
  \mnras, 461, 3683

\bibitem[{{Izotov} {et~al.}(2018{\natexlab{a}}){Izotov}, {Schaerer}, {Worseck},
  {Guseva}, {Thuan}, {Verhamme}, {Orlitov{\'a}}, \& {Fricke}}]{izotov2018a}
{Izotov}, Y.~I., {Schaerer}, D., {Worseck}, G., {et~al.} 2018{\natexlab{a}},
  \mnras, 474, 4514

\bibitem[{{Izotov} {et~al.}(2018{\natexlab{b}}){Izotov}, {Worseck}, {Schaerer},
  {Guseva}, {Thuan}, {Fricke}, \& {Orlitov{\'a}}}]{izotov2018b}
{Izotov}, Y.~I., {Worseck}, G., {Schaerer}, D., {et~al.} 2018{\natexlab{b}},
  \mnras, 478, 4851

\bibitem[{{Jaskot} {et~al.}(2019){Jaskot}, {Dowd}, {Oey}, {Scarlata}, \&
  {McKinney}}]{jaskot2019}
{Jaskot}, A.~E., {Dowd}, T., {Oey}, M.~S., {Scarlata}, C., \& {McKinney}, J.
  2019, \apj, 885, 96

\bibitem[{{Jaskot} \& {Oey}(2013)}]{jaskot2013}
{Jaskot}, A.~E. \& {Oey}, M.~S. 2013, \apj, 766, 91

\bibitem[{{Ji} {et~al.}(2020){Ji}, {Giavalisco}, {Vanzella}, {Siana},
  {Pentericci}, {Jaskot}, {Liu}, {Nonino}, {Ferguson}, {Castellano},
  {Mannucci}, {Schaerer}, {Fynbo}, {Papovich}, {Carnall}, {Amorin}, {Simons},
  {Hathi}, {Cullen}, \& {McLeod}}]{ji2020}
{Ji}, Z., {Giavalisco}, M., {Vanzella}, E., {et~al.} 2020, \apj, 888, 109

\bibitem[{{Jones} {et~al.}(2021){Jones}, {Barger}, \& {Cowie}}]{jones2021}
{Jones}, L.~H., {Barger}, A.~J., \& {Cowie}, L.~L. 2021, \apj, 908, 222

\bibitem[{{Kakiichi} \& {Gronke}(2021)}]{kakiichi2021}
{Kakiichi}, K. \& {Gronke}, M. 2021, \apj, 908, 30

\bibitem[{{Keenan} {et~al.}(2017){Keenan}, {Oey}, {Jaskot}, \&
  {James}}]{keenan2017}
{Keenan}, R.~P., {Oey}, M.~S., {Jaskot}, A.~E., \& {James}, B.~L. 2017, \apj,
  848, 12

\bibitem[{{Koekemoer} {et~al.}(2013){Koekemoer}, {Ellis}, {McLure}, {Dunlop},
  {Robertson}, {Ono}, {Schenker}, {Ouchi}, {Bowler}, {Rogers}, {Curtis-Lake},
  {Schneider}, {Charlot}, {Stark}, {Furlanetto}, {Cirasuolo}, {Wild}, \&
  {Targett}}]{koekemoer2013}
{Koekemoer}, A.~M., {Ellis}, R.~S., {McLure}, R.~J., {et~al.} 2013, \apjs, 209,
  3

\bibitem[{{Leitet} {et~al.}(2013){Leitet}, {Bergvall}, {Hayes}, {Linn{\'e}}, \&
  {Zackrisson}}]{leitet2013}
{Leitet}, E., {Bergvall}, N., {Hayes}, M., {Linn{\'e}}, S., \& {Zackrisson}, E.
  2013, \aap, 553, A106

\bibitem[{{Leitet} {et~al.}(2011){Leitet}, {Bergvall}, {Piskunov}, \&
  {Andersson}}]{leitet2011}
{Leitet}, E., {Bergvall}, N., {Piskunov}, N., \& {Andersson}, B.-G. 2011, \aap,
  532, A107

\bibitem[{{Leitherer} {et~al.}(2016){Leitherer}, {Hernandez}, {Lee}, \&
  {Oey}}]{leitherer2016}
{Leitherer}, C., {Hernandez}, S., {Lee}, J.~C., \& {Oey}, M.~S. 2016, \apj,
  823, 64

\bibitem[{{Madau} \& {Dickinson}(2014)}]{madau2014}
{Madau}, P. \& {Dickinson}, M. 2014, \araa, 52, 415

\bibitem[{{Madau} \& {Haardt}(2015)}]{madau2015}
{Madau}, P. \& {Haardt}, F. 2015, \apjl, 813, L8

\bibitem[{{Malkan} \& {Malkan}(2021)}]{malkan2021b}
{Malkan}, M.~A. \& {Malkan}, B.~K. 2021, \apj, 909, 92

\bibitem[{{Marques-Chaves} {et~al.}(2021){Marques-Chaves}, {Schaerer},
  {{\'A}lvarez-M{\'a}rquez}, {Colina}, {Dessauges-Zavadsky},
  {P{\'e}rez-Fournon}, {Saldana-Lopez}, \& {Verhamme}}]{marqueschaves2021}
{Marques-Chaves}, R., {Schaerer}, D., {{\'A}lvarez-M{\'a}rquez}, J., {et~al.}
  2021, \mnras, 507, 524

\bibitem[{{Matthee} {et~al.}(2022){Matthee}, {Naidu}, {Pezzulli}, {Gronke},
  {Sobral}, {Oesch}, {Hayes}, {Erb}, {Schaerer}, {Amor{\'\i}n}, {Tacchella},
  {Paulino-Afonso}, {Llerena}, {Calhau}, \& {R{\"o}ttgering}}]{matthee2022}
{Matthee}, J., {Naidu}, R.~P., {Pezzulli}, G., {et~al.} 2022, \mnras, 512, 5960

\bibitem[{{McLure} {et~al.}(2018){McLure}, {Pentericci}, {Cimatti}, {Dunlop},
  {Elbaz}, {Fontana}, {Nandra}, {Amorin}, {Bolzonella}, {Bongiorno}, {Carnall},
  {Castellano}, {Cirasuolo}, {Cucciati}, {Cullen}, {De Barros}, {Finkelstein},
  {Fontanot}, {Franzetti}, {Fumana}, {Gargiulo}, {Garilli}, {Guaita},
  {Hartley}, {Iovino}, {Jarvis}, {Juneau}, {Karman}, {Maccagni}, {Marchi},
  {M{\'a}rmol-Queralt{\'o}}, {Pompei}, {Pozzetti}, {Scodeggio}, {Sommariva},
  {Talia}, {Almaini}, {Balestra}, {Bardelli}, {Bell}, {Bourne}, {Bowler},
  {Brusa}, {Buitrago}, {Caputi}, {Cassata}, {Charlot}, {Citro}, {Cresci},
  {Cristiani}, {Curtis-Lake}, {Dickinson}, {Fazio}, {Ferguson}, {Fiore},
  {Franco}, {Fynbo}, {Galametz}, {Georgakakis}, {Giavalisco}, {Grazian},
  {Hathi}, {Jung}, {Kim}, {Koekemoer}, {Khusanova}, {Le F{\`e}vre}, {Lotz},
  {Mannucci}, {Maltby}, {Matsuoka}, {McLeod}, {Mendez-Hernandez},
  {Mendez-Abreu}, {Mignoli}, {Moresco}, {Mortlock}, {Nonino}, {Pannella},
  {Papovich}, {Popesso}, {Rosario}, {Salvato}, {Santini}, {Schaerer},
  {Schreiber}, {Stark}, {Tasca}, {Thomas}, {Treu}, {Vanzella}, {Wild},
  {Williams}, {Zamorani}, \& {Zucca}}]{mclure2018}
{McLure}, R.~J., {Pentericci}, L., {Cimatti}, A., {et~al.} 2018, \mnras, 479,
  25

\bibitem[{{Menacho} {et~al.}(2021){Menacho}, {{\"O}stlin}, {Bik}, {Adamo},
  {Bergvall}, {Della Bruna}, {Hayes}, {Melinder}, \&
  {Rivera-Thorsen}}]{menacho2021}
{Menacho}, V., {{\"O}stlin}, G., {Bik}, A., {et~al.} 2021, \mnras, 506, 1777

\bibitem[{{Menacho} {et~al.}(2019){Menacho}, {{\"O}stlin}, {Bik}, {Della
  Bruna}, {Melinder}, {Adamo}, {Hayes}, {Herenz}, \& {Bergvall}}]{menacho2019}
{Menacho}, V., {{\"O}stlin}, G., {Bik}, A., {et~al.} 2019, \mnras, 487, 3183

\bibitem[{{Mostardi} {et~al.}(2015){Mostardi}, {Shapley}, {Steidel}, {Trainor},
  {Reddy}, \& {Siana}}]{mostardi2015}
{Mostardi}, R.~E., {Shapley}, A.~E., {Steidel}, C.~C., {et~al.} 2015, \apj,
  810, 107

\bibitem[{{Moster} {et~al.}(2011){Moster}, {Somerville}, {Newman}, \&
  {Rix}}]{moster2011}
{Moster}, B.~P., {Somerville}, R.~S., {Newman}, J.~A., \& {Rix}, H.-W. 2011,
  \apj, 731, 113

\bibitem[{{Naidu} {et~al.}(2022){Naidu}, {Matthee}, {Oesch}, {Conroy},
  {Sobral}, {Pezzulli}, {Hayes}, {Erb}, {Amor{\'\i}n}, {Gronke}, {Schaerer},
  {Tacchella}, {Kerutt}, {Paulino-Afonso}, {Calhau}, {Llerena}, \&
  {R{\"o}ttgering}}]{naidu2022}
{Naidu}, R.~P., {Matthee}, J., {Oesch}, P.~A., {et~al.} 2022, \mnras, 510, 4582

\bibitem[{{Naidu} {et~al.}(2020){Naidu}, {Tacchella}, {Mason}, {Bose}, {Oesch},
  \& {Conroy}}]{naidu2020a}
{Naidu}, R.~P., {Tacchella}, S., {Mason}, C.~A., {et~al.} 2020, \apj, 892, 109

\bibitem[{{Nakajima} {et~al.}(2020){Nakajima}, {Ellis}, {Robertson}, {Tang}, \&
  {Stark}}]{nakajima2020}
{Nakajima}, K., {Ellis}, R.~S., {Robertson}, B.~E., {Tang}, M., \& {Stark},
  D.~P. 2020, \apj, 889, 161

\bibitem[{{Nakajima} \& {Ouchi}(2014)}]{nakajima2014}
{Nakajima}, K. \& {Ouchi}, M. 2014, \mnras, 442, 900

\bibitem[{{Nestor} {et~al.}(2013){Nestor}, {Shapley}, {Kornei}, {Steidel}, \&
  {Siana}}]{nestor2013}
{Nestor}, D.~B., {Shapley}, A.~E., {Kornei}, K.~A., {Steidel}, C.~C., \&
  {Siana}, B. 2013, \apj, 765, 47

\bibitem[{{Newman} \& {Moster}(2014)}]{newman2014}
{Newman}, J.~A. \& {Moster}, B.~P. 2014, {QUICKCV: Cosmic variance calculator}

\bibitem[{{Oesch} {et~al.}(2010){Oesch}, {Bouwens}, {Carollo}, {Illingworth},
  {Trenti}, {Stiavelli}, {Magee}, {Labb{\'e}}, \& {Franx}}]{oesch2010b}
{Oesch}, P.~A., {Bouwens}, R.~J., {Carollo}, C.~M., {et~al.} 2010, \apjl, 709,
  L21

\bibitem[{{{\"O}stlin} {et~al.}(2015){{\"O}stlin}, {Marquart}, {Cumming},
  {Fathi}, {Bergvall}, {Adamo}, {Amram}, \& {Hayes}}]{ostlin2015}
{{\"O}stlin}, G., {Marquart}, T., {Cumming}, R.~J., {et~al.} 2015, \aap, 583,
  A55

\bibitem[{{{\"O}stlin} {et~al.}(2021){{\"O}stlin}, {Rivera-Thorsen}, {Menacho},
  {Hayes}, {Runnholm}, {Micheva}, {Oey}, {Adamo}, {Bik}, {Cannon}, {Gronke},
  {Kunth}, {Laursen}, {Mas-Hesse}, {Melinder}, {Messa}, {Sirressi}, \&
  {Smith}}]{ostlin2021}
{{\"O}stlin}, G., {Rivera-Thorsen}, T.~E., {Menacho}, V., {et~al.} 2021, \apj,
  912, 155

\bibitem[{{Puschnig} {et~al.}(2017){Puschnig}, {Hayes}, {{\"O}stlin},
  {Rivera-Thorsen}, {Melinder}, {Cannon}, {Menacho}, {Zackrisson}, {Bergvall},
  \& {Leitet}}]{puschnig2017}
{Puschnig}, J., {Hayes}, M., {{\"O}stlin}, G., {et~al.} 2017, \mnras, 469, 3252

\bibitem[{Rafelski {et~al.}(2015)Rafelski, Teplitz, Gardner, Coe, Bond,
  Koekemoer, Grogin, Kurczynski, McGrath, Bourque, Atek, Brown, Colbert,
  Codoreanu, Ferguson, Finkelstein, Gawiser, Giavalisco, Gronwall, Hanish, Lee,
  Mehta, de~Mello, Ravindranath, Ryan, Scarlata, Siana, Soto, \&
  Voyer}]{rafelski15_uvudf}
Rafelski, M., Teplitz, H.~I., Gardner, J.~P., {et~al.} 2015, The Astronomical
  Journal, 150, 31

\bibitem[{{Ramachandran} {et~al.}(2018{\natexlab{a}}){Ramachandran}, {Hainich},
  {Hamann}, {Oskinova}, {Shenar}, {Sander}, {Todt}, \&
  {Gallagher}}]{ramachandran2018a}
{Ramachandran}, V., {Hainich}, R., {Hamann}, W.~R., {et~al.}
  2018{\natexlab{a}}, \aap, 609, A7

\bibitem[{{Ramachandran} {et~al.}(2018{\natexlab{b}}){Ramachandran}, {Hamann},
  {Hainich}, {Oskinova}, {Shenar}, {Sander}, {Todt}, \&
  {Gallagher}}]{ramachandran2018}
{Ramachandran}, V., {Hamann}, W.-R., {Hainich}, R., {et~al.}
  2018{\natexlab{b}}, \aap, 615, A40

\bibitem[{{Ramachandran} {et~al.}(2018{\natexlab{c}}){Ramachandran}, {Hamann},
  {Hainich}, {Oskinova}, {Shenar}, {Sander}, {Todt}, \&
  {Gallagher}}]{ramachandran2018b}
{Ramachandran}, V., {Hamann}, W.~R., {Hainich}, R., {et~al.}
  2018{\natexlab{c}}, \aap, 615, A40

\bibitem[{{Ramachandran} {et~al.}(2019){Ramachandran}, {Hamann}, {Oskinova},
  {Gallagher}, {Hainich}, {Shenar}, {Sander}, {Todt}, \&
  {Fulmer}}]{ramachandran2019}
{Ramachandran}, V., {Hamann}, W.~R., {Oskinova}, L.~M., {et~al.} 2019, \aap,
  625, A104

\bibitem[{{Ramachandran} {et~al.}(2021){Ramachandran}, {Oskinova}, \&
  {Hamann}}]{ramachandran2021}
{Ramachandran}, V., {Oskinova}, L.~M., \& {Hamann}, W.~R. 2021, \aap, 646, A16

\bibitem[{{Reddy} {et~al.}(2016){Reddy}, {Steidel}, {Pettini}, \&
  {Bogosavljevic}}]{reddy2016}
{Reddy}, N.~A., {Steidel}, C.~C., {Pettini}, M., \& {Bogosavljevic}, M. 2016,
  \apj, 828, 107

\bibitem[{{Rivera-Thorsen} {et~al.}(2019){Rivera-Thorsen}, {Dahle}, {Chisholm},
  {Florian}, {Gronke}, {Rigby}, {Gladders}, {Mahler}, {Sharon}, \&
  {Bayliss}}]{riverathorsen2019}
{Rivera-Thorsen}, T.~E., {Dahle}, H., {Chisholm}, J., {et~al.} 2019, Science,
  366, 738

\bibitem[{{Robertson} {et~al.}(2015){Robertson}, {Ellis}, {Furlanetto}, \&
  {Dunlop}}]{robertson2015}
{Robertson}, B.~E., {Ellis}, R.~S., {Furlanetto}, S.~R., \& {Dunlop}, J.~S.
  2015, \apjl, 802, L19

\bibitem[{{Rutkowski} {et~al.}(2016){Rutkowski}, {Scarlata}, {Haardt}, {Siana},
  {Henry}, {Rafelski}, {Hayes}, {Salvato}, {Pahl}, {Mehta}, {Beck}, {Malkan},
  \& {Teplitz}}]{rutkowski2016}
{Rutkowski}, M.~J., {Scarlata}, C., {Haardt}, F., {et~al.} 2016, \apj, 819, 81

\bibitem[{{Rutkowski} {et~al.}(2017){Rutkowski}, {Scarlata}, {Henry}, {Hayes},
  {Mehta}, {Hathi}, {Cohen}, {Windhorst}, {Koekemoer}, {Teplitz}, {Haardt}, \&
  {Siana}}]{rutkowski2017}
{Rutkowski}, M.~J., {Scarlata}, C., {Henry}, A., {et~al.} 2017, \apjl, 841, L27

\bibitem[{{Saha} {et~al.}(2020){Saha}, {Tandon}, {Simmonds}, {Verhamme},
  {Paswan}, {Schaerer}, {Rutkowski}, {Borgohain}, {Elmegreen}, {Inoue},
  {Combes}, {Elmegreen}, \& {Paalvast}}]{saha2020}
{Saha}, K., {Tandon}, S.~N., {Simmonds}, C., {et~al.} 2020, Nature Astronomy
  [\eprint[arXiv]{2008.11394}]

\bibitem[{{Saxena} {et~al.}(2022){Saxena}, {Pentericci}, {Ellis}, {Guaita},
  {Calabr{\`o}}, {Schaerer}, {Vanzella}, {Amor{\'\i}n}, {Bolzonella},
  {Castellano}, {Fontanot}, {Hathi}, {Hibon}, {Llerena}, {Mannucci},
  {Saldana-Lopez}, {Talia}, \& {Zamorani}}]{saxena2021}
{Saxena}, A., {Pentericci}, L., {Ellis}, R.~S., {et~al.} 2022, \mnras, 511, 120

\bibitem[{{Shapley} {et~al.}(2003){Shapley}, {Steidel}, {Pettini}, \&
  {Adelberger}}]{shapley2003}
{Shapley}, A.~E., {Steidel}, C.~C., {Pettini}, M., \& {Adelberger}, K.~L. 2003,
  \apj, 588, 65

\bibitem[{{Shapley} {et~al.}(2016){Shapley}, {Steidel}, {Strom},
  {Bogosavljevi{\'c}}, {Reddy}, {Siana}, {Mostardi}, \& {Rudie}}]{shapley2016}
{Shapley}, A.~E., {Steidel}, C.~C., {Strom}, A.~L., {et~al.} 2016, \apjl, 826,
  L24

\bibitem[{{Siana} {et~al.}(2007){Siana}, {Teplitz}, {Colbert}, {Ferguson},
  {Dickinson}, {Brown}, {Conselice}, {de Mello}, {Gardner}, {Giavalisco}, \&
  {Menanteau}}]{siana2007}
{Siana}, B., {Teplitz}, H.~I., {Colbert}, J., {et~al.} 2007, \apj, 668, 62

\bibitem[{{Siana} {et~al.}(2010){Siana}, {Teplitz}, {Ferguson}, {Brown},
  {Giavalisco}, {Dickinson}, {Chary}, {de Mello}, {Conselice}, {Bridge},
  {Gardner}, {Colbert}, \& {Scarlata}}]{siana2010}
{Siana}, B., {Teplitz}, H.~I., {Ferguson}, H.~C., {et~al.} 2010, \apj, 723, 241

\bibitem[{{Stark} {et~al.}(2010){Stark}, {Ellis}, {Chiu}, {Ouchi}, \&
  {Bunker}}]{stark2010}
{Stark}, D.~P., {Ellis}, R.~S., {Chiu}, K., {Ouchi}, M., \& {Bunker}, A. 2010,
  \mnras, 408, 1628

\bibitem[{{Steidel} {et~al.}(2018){Steidel}, {Bogosavljevi{\'c}}, {Shapley},
  {Reddy}, {Rudie}, {Pettini}, {Trainor}, \& {Strom}}]{steidel2018}
{Steidel}, C.~C., {Bogosavljevi{\'c}}, M., {Shapley}, A.~E., {et~al.} 2018,
  \apj, 869, 123

\bibitem[{{Steidel} {et~al.}(2001){Steidel}, {Pettini}, \&
  {Adelberger}}]{steidel2001}
{Steidel}, C.~C., {Pettini}, M., \& {Adelberger}, K.~L. 2001, \apj, 546, 665

\bibitem[{{STScI Development Team}(2013)}]{pysynphot2013}
{STScI Development Team}. 2013, {pysynphot: Synthetic photometry software
  package}, Astrophysics Source Code Library, record ascl:1303.023

\bibitem[{{Teplitz} {et~al.}(2013){Teplitz}, {Rafelski}, {Kurczynski}, {Bond},
  {Grogin}, {Koekemoer}, {Atek}, {Brown}, {Coe}, {Colbert}, {Ferguson},
  {Finkelstein}, {Gardner}, {Gawiser}, {Giavalisco}, {Gronwall}, {Hanish},
  {Lee}, {de Mello}, {Ravindranath}, {Ryan}, {Siana}, {Scarlata}, {Soto},
  {Voyer}, \& {Wolfe}}]{teplitz2013}
{Teplitz}, H.~I., {Rafelski}, M., {Kurczynski}, P., {et~al.} 2013, \aj, 146,
  159

\bibitem[{{Trenti} \& {Stiavelli}(2008)}]{trenti2008}
{Trenti}, M. \& {Stiavelli}, M. 2008, \apj, 676, 767

\bibitem[{{Vanzella} {et~al.}(2015){Vanzella}, {de Barros}, {Castellano},
  {Grazian}, {Inoue}, {Schaerer}, {Guaita}, {Zamorani}, {Giavalisco}, {Siana},
  {Pentericci}, {Giallongo}, {Fontana}, \& {Vignali}}]{vanzella2015}
{Vanzella}, E., {de Barros}, S., {Castellano}, M., {et~al.} 2015, \aap, 576,
  A116

\bibitem[{{Vanzella} {et~al.}(2016){Vanzella}, {de Barros}, {Vasei}, {Alavi},
  {Giavalisco}, {Siana}, {Grazian}, {Hasinger}, {Suh}, {Cappelluti}, {Vito},
  {Amorin}, {Balestra}, {Brusa}, {Calura}, {Castellano}, {Comastri}, {Fontana},
  {Gilli}, {Mignoli}, {Pentericci}, {Vignali}, \& {Zamorani}}]{vanzella2016}
{Vanzella}, E., {de Barros}, S., {Vasei}, K., {et~al.} 2016, \apj, 825, 41

\bibitem[{{Vanzella} {et~al.}(2012){Vanzella}, {Guo}, {Giavalisco}, {Grazian},
  {Castellano}, {Cristiani}, {Dickinson}, {Fontana}, {Nonino}, {Giallongo},
  {Pentericci}, {Galametz}, {Faber}, {Ferguson}, {Grogin}, {Koekemoer},
  {Newman}, \& {Siana}}]{vanzella2012}
{Vanzella}, E., {Guo}, Y., {Giavalisco}, M., {et~al.} 2012, \apj, 751, 70

\bibitem[{{Vanzella} {et~al.}(2018){Vanzella}, {Nonino}, {Cupani},
  {Castellano}, {Sani}, {Mignoli}, {Calura}, {Meneghetti}, {Gilli}, {Comastri},
  {Mercurio}, {Caminha}, {Caputi}, {Rosati}, {Grillo}, {Cristiani}, {Balestra},
  {Fontana}, \& {Giavalisco}}]{vanzella2018}
{Vanzella}, E., {Nonino}, M., {Cupani}, G., {et~al.} 2018, \mnras, 476, L15

\bibitem[{{Vasei} {et~al.}(2016){Vasei}, {Siana}, {Shapley}, {Quider}, {Alavi},
  {Rafelski}, {Steidel}, {Pettini}, \& {Lewis}}]{vasei2016}
{Vasei}, K., {Siana}, B., {Shapley}, A.~E., {et~al.} 2016, \apj, 831, 38

\bibitem[{{Verhamme} {et~al.}(2015){Verhamme}, {Orlitov{\'a}}, {Schaerer}, \&
  {Hayes}}]{verhamme2015}
{Verhamme}, A., {Orlitov{\'a}}, I., {Schaerer}, D., \& {Hayes}, M. 2015, \aap,
  578, A7

\bibitem[{{Wang} {et~al.}(2019){Wang}, {Heckman}, {Leitherer}, {Alexandroff},
  {Borthakur}, \& {Overzier}}]{wang2019}
{Wang}, B., {Heckman}, T.~M., {Leitherer}, C., {et~al.} 2019, \apj, 885, 57

\end{thebibliography}

\begin{appendix}
\section{Background subtraction                              \label{app.bgsub}}
\label{sec:org19d5ed8}
For the sake of reproducibility, we tabulate in Table~\ref{tab:bgsubtpars} for each
filter the box size passed to the \texttt{photutils.Background2D} class and the each of
the filters, as well as the values of \texttt{nsigma}, \texttt{npixels} and \texttt{filter\_fwhm} passed to
the \texttt{photutils.make\_source\_mask} routine used for background subtraction. All
other parameters are kept at their default value.

\begin{table}[tbhp]
\caption{\label{tab:bgsubtpars}Parameters for background subtraction and source mask creation listed for each filter.}
\centering
\begin{tabular}{lrrrr}
\toprule
Filter & Box size & SNR & Npixels & FWHM\\
\midrule
\midrule
F225W & 35 & 5 & 9 & 3\\
F275W & 35 & 5 & 25 & 5\\
F336W & 65 & 5 & 25 & 5\\
F435W & 45 & 5 & 25 & 5\\
F606W & 45 & 5 & 36 & 5\\
F775W & 45 & 5 & 49 & 5\\
F850LP & 45 & 10 & 64 & 5\\
F105W & 75 & 75 & 49 & 5\\
F125W & 75 & 75 & 49 & 5\\
F140W & 75 & 75 & 49 & 5\\
F160W & 75 & 75 & 49 & 5\\
\bottomrule
\end{tabular}
\end{table}

\begin{figure*}[tbhp]
\centering
\includegraphics[width=.75\textwidth]{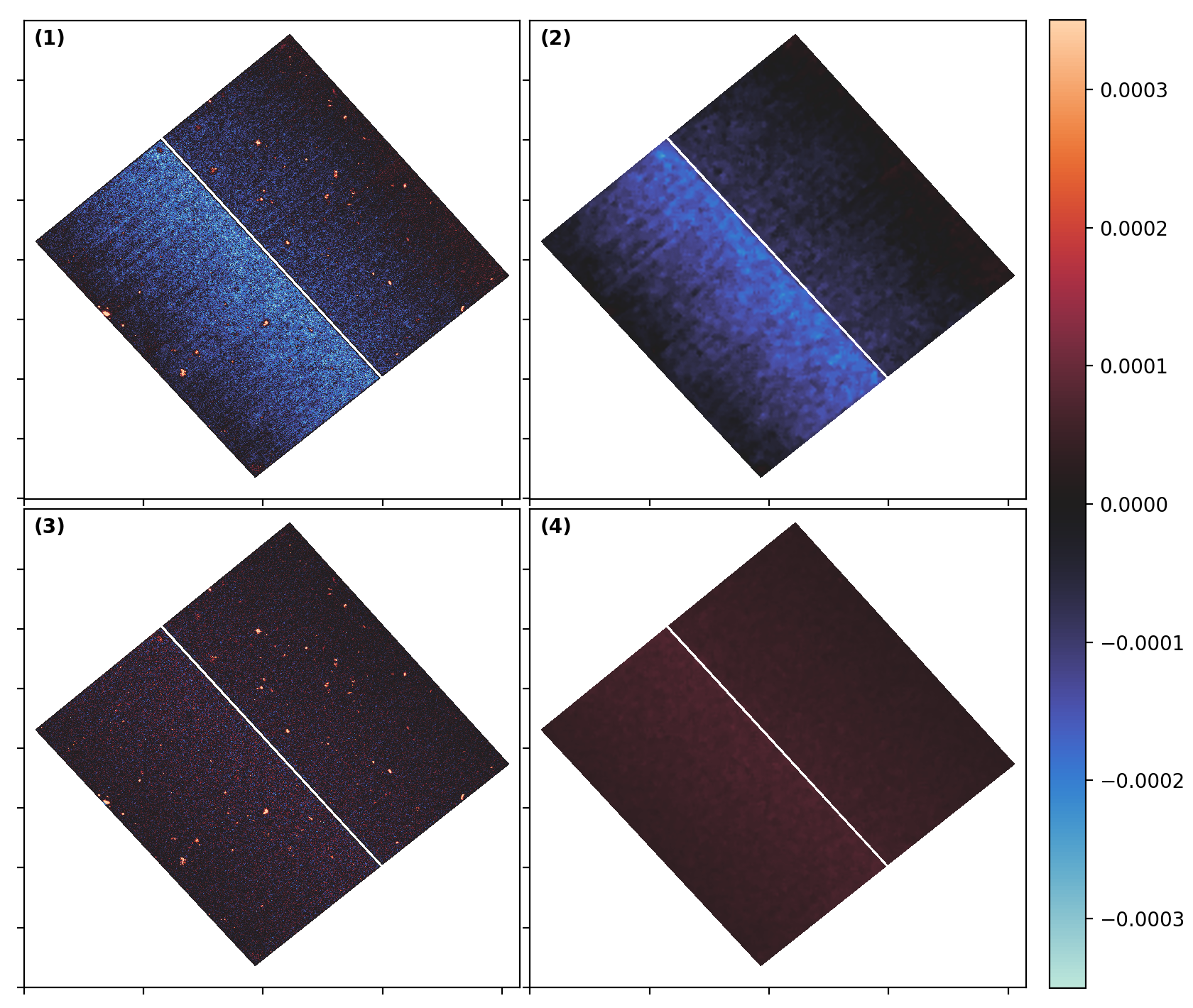}
\caption{\label{fig:bgsub225}Background subtraction of the PSF matched F225W frame after the second iteration. \textbf{(1)}: Raw data. \textbf{(2)}: Computed background. \textbf{(3)}: Background-subtracted data. \textbf{(4)}: Uncertainties on the background. All panels share the same color cuts, selected to emphasize structure in the background.}
\end{figure*}

\begin{figure*}[htb]
\centering
\includegraphics[width=.75\textwidth]{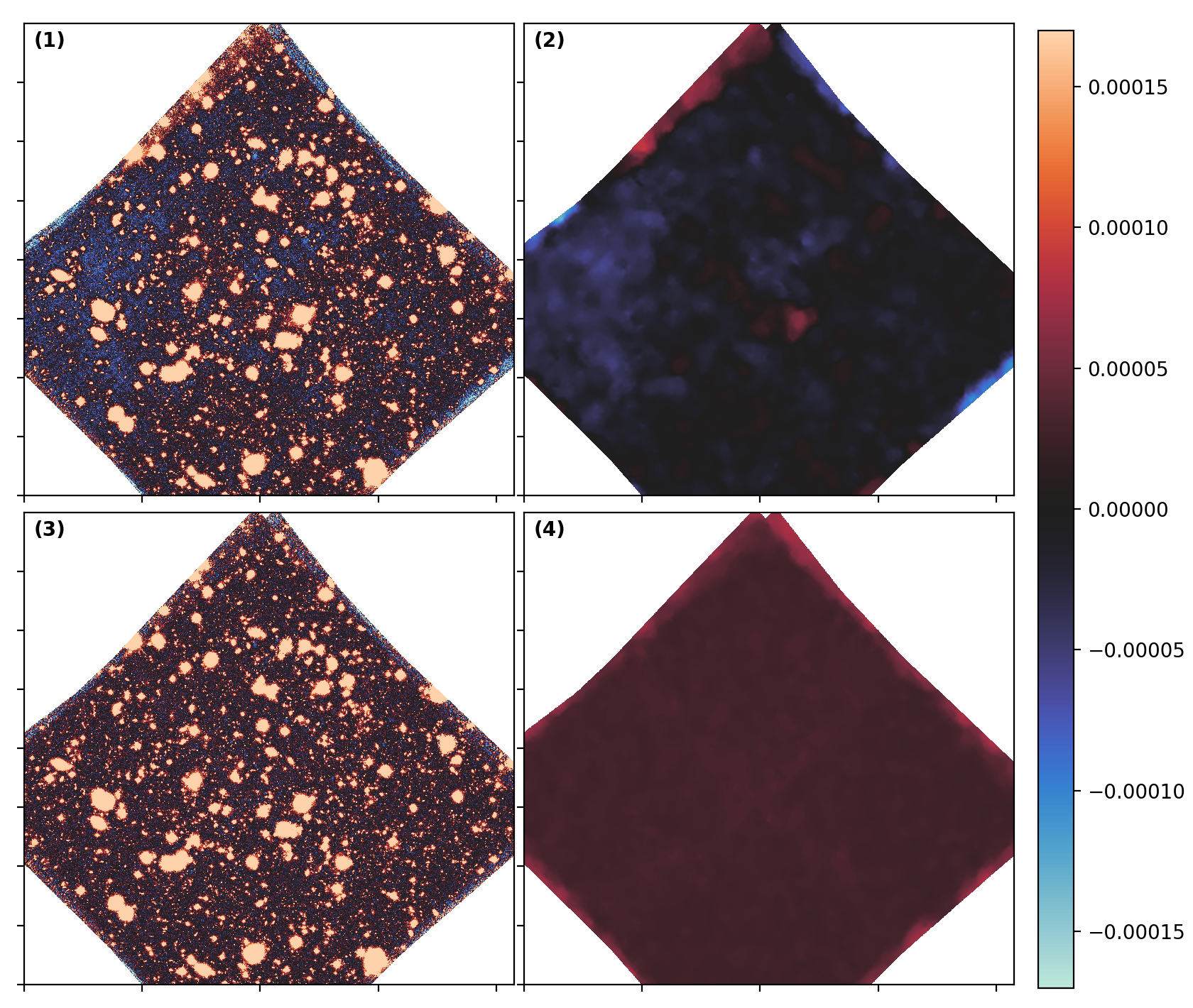}
\caption{\label{fig:bgsub775}Example background subtraction of the PSF matched F775W frame. Panel content is the same as in fig. \ref{fig:bgsub225}, except the upper and lower color cuts are half those of the fig. \ref{fig:bgsub225}.}
\end{figure*}

Figs.~\ref{fig:bgsub225}  and \ref{fig:bgsub775}  show examples  of some  of the
challenges associated with background subtraction in UV  and optical/red bands.
Both show in panel (1) the PSF-matched data  including the  background, in
panel (2)  the modeled background, in panel (3) the background subtracted
data, and in panel (4) the uncertainties  on the background.
Fig.~\ref{fig:bgsub225} shows  these for the F225W observations.  The image  in
(1)  is characterized  by few  and relatively faint  sources and  a highly
structured  background, with  stripes and  spatial variations in the level. The
background levels are numerically comparable to the flux  in many  of  the
fainter objects,  and  to  achieve reliable  photometric results, it is
necessary to model  and subtract this structure, without modeling and
subtracting any actual sources. The modeling is detailed in
sect.~\ref{sec:orgf584df0}.

Fig.~\ref{fig:bgsub775}  shows the  same  process,  but for  the  F775W filter.  Here,
sources are more  numerous and much brighter in comparison  with the background.
Additionally, giant elliptical galaxies with their old, red and UV-faint stellar
populations  are  becoming  visible  in  these  bands.  These  galaxies  have  a
substantially larger extent than the UV  sources in F225W above, meaning that to
avoid  modeling and  subtracting them,  either a  larger kernel  or wider  masks
around  detected sources  are necessary.  But these  masks cannot  be too  wide,
either, as the large number of small  and faint sources in the field would leave
very few pixels with which to model the background.

Panel (2)  in fig.~\ref{fig:bgsub775} shows  one such  giant elliptical which  has not
been entirely masked out, such that  the low surface brightness outer parts have
been modeled and subtracted. This was  the best compromise we could find between
masking and leaving pixels to model without resorting to manual masking. We have
checked if any of our sources are  affected by this (weak) effect, which was not
the case.

\section{Source detection and deblending settings}
\label{sec:orgd944c04}
\subsection{Source detection settings}
\label{sec:org41467ed}
For each of the UV filters F225W, F275W and F336W, we ran the \texttt{photutils} source
detection function \texttt{detect\_sources} with a set of parameters selected to find all
sources above a set of threshold values of signal-to-noise, Due to the
comparatively low SNR in the UV filters, and the blotchy noise patterns
described above, it can be difficult to distinguish spurious detections from
real sources in each individual detection filter. At later steps, however, we
have imaging data in up to 10 additional filters available, meaning that we have
opted for a set of very permissible source finding and detection settings,
deliberately resulting in a large number of spurious detections which we then
subsequently reject using the additional information provided by the additional
data. In order to not miss any real but faint objects, we deliberately selected
the detection parameters to allow for spurious detections. These could then be
filtered out in subsequent steps based on optical and IR photometry in the
segmentation maps. We found a set of suitable parameters for source detection
and deblending by trial and error, eventually settling on the values tabulated
in Table~\ref{tab:detdeb}.

\subsection{Deblending}
\label{sec:orge0e6dbc}
In deblending autodetected sources, one must strike a balance between on one
hand separating truly distinct sources, while on the other hand correctly
identifying substructure within a galaxy as such. The main purpose of deblending
in the context of this work has been to correctly identify as many sources as
possible at redshifts 2-4. We have therefore, as with source detection, pursued
a quite aggressive deblending strategy, where correctly identifying unresolved
or barely resolved but potentially overlapping sources took priority. This
resulted in a number of clumps in large spirals identified as separate sources,
but these were easily identified and rejected at later steps (see
Sect.~\ref{sec:filtering}, especially \ref{sec:morphvis}).

A different challenge was posed by the blotchy pattern described in sect.~\ref{sec:orgf584df0}. These patterns are sometimes difficult to tell apart
from lower surface brightness extended features, and thus a balance needed to be
found between on one side recovering all flux from true physical sources, while
on the other hand avoiding to include too extended blotches in the segmentation
maps.

Parameters for source detection and deblending were set in a process of trial
and error, and the final choices are presented in Table~\ref{tab:detdeb}.

In  table  \ref{tab:detdeb}  we  tabulate  parameters  passed  to  the  \texttt{photutils}  functions
\texttt{detect\_sources} and  \texttt{deblend\_sources}. Most are identical  for the three filters  but listed
for completeness and easy access. The process is described in sect.~\ref{sec:orge0e6dbc}. FWHM is
the width in pixels of the Gaussian convolution kernel used for deblending; npixels is the
minimum number of consecutive pixels required to be above the threshold SNR. Nlevels and
contrast are explained in the \texttt{photutils} documentation.

\begin{table}[tbhp]
\caption{\label{tab:detdeb}Detection and deblending parameters used in the three detection filters.}
\centering
\begin{tabular}{lrrrrr}
\toprule
Det. filter & npixels & snr & fwhm & nlevels & contrast\\
\midrule
F225W & 16 & 3 & 5 & 32 & 0.05\\
F275W & 16 & 3 & 5 & 32 & 0.05\\
F336W & 25 & 3 & 5 & 32 & 0.03\\
\bottomrule
\end{tabular}
\end{table}

\subsection{How many spurious detections?                   \label{sec:spurious}}
\label{sec:orgc60341e}
When detecting sources, we have deliberately set the detection criteria to be
permissive, allowing for a relatively large number of spurious detection and
relying on later filtering steps to discern these from real objects. This of
course raises a concern regarding how effective these filtering steps are. With
sufficiently many spurious detections present, the probability grows that some
of these might by slip through the filtering mechanisms set up to catch them.

We tested the magnitude of this problem by inverting the sign of the pixels in
the background-subtracted F225W detection frame, and running the source
detection routine with the exact same settings as before. By giving all true
sources negative flux, we ensured that all detections were due to noise. We then
proceeded as previously, by extracting flux from the neighboring filter on the
red side, in this case F275W, in the segments created by the detection routine,
and comparing colors and SNR as described for the science frames above, to test
how large a fraction of the initial detections made it past these filtering
criteria.

In the inverted F225W frame, we found 2317 ``sources'' living up to our detection
criteria. After extracting the corresponding fluxes in F275W and comparing SNR
and demanding that the AB magnitude be larger in the bluer band, we found that
18 of these sources still remained. We then tested these against the R15 and I17
catalogs with the same angular distance requirement of \(0 \farcs 75\) as above; 5
of the 18 sources still remaining fell inside this radius of a catalog object,
yielding a spurious detection rate of 5/2317, or 0.2\%, before visual inspection.

\section{UV filter error estimates}\label{sec:org8bf15ba}
We ran the source detection routines in all the three UV filters. For each of
these, we measured fluxes for every segment in the resulting segmentation
filters in all 11 available filters. Visual inspection of a number of sources
made us suspect that the errors in the UV filters were over estimated, which we
confirmed by running Monte Carlo sampling, as described in the main text.

\begin{figure*}[tbh]
\centering
\includegraphics[width=0.9\textwidth]{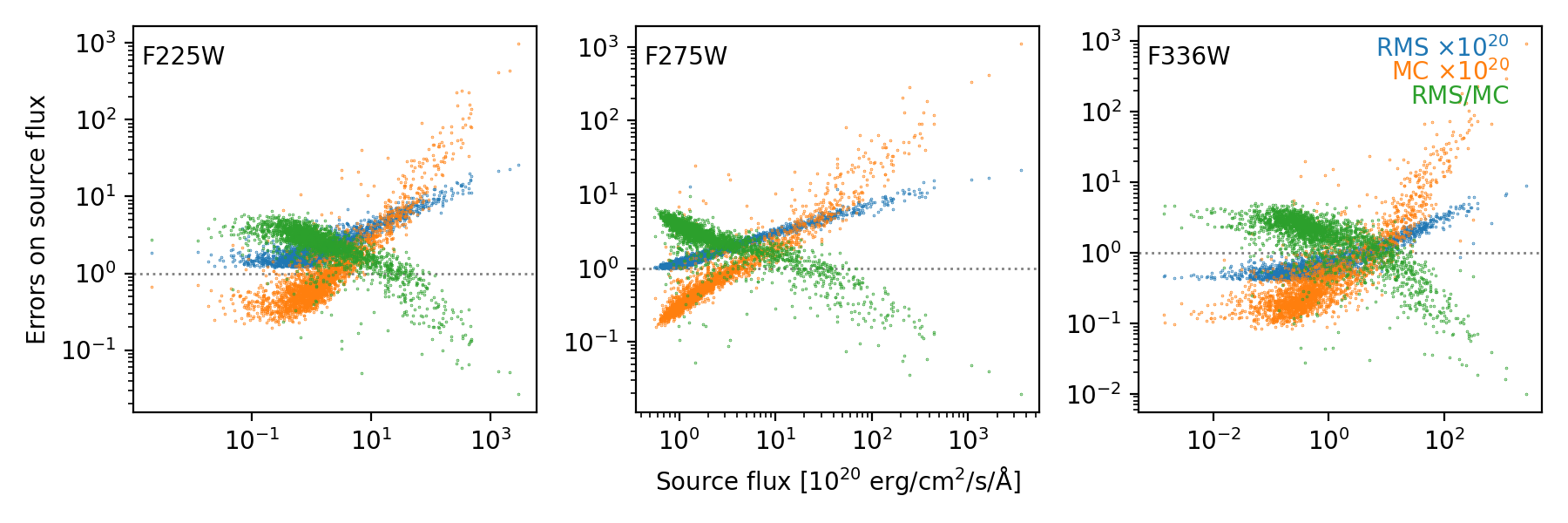}
\caption{\label{fig:errors}RMS map based (blue) and MC computed (orange) errors (\(\times 10^{20}\)) plotted vs.\ aperture fluxes for the source catalog detected using F275W. The ratio between the two is shown in green for easier comparison.}
\end{figure*}

In fig.~\ref{fig:errors}, we show as an example the originally measured and the
re-estimated flux errors from the catalog detected using F275W. Objects with
source sums \(\lesssim\) 10\textsuperscript{-3}, and typical SNR of \(\sim\) 1, are spurious detections in
F275W and are filtered out of our sample at the next step.

At the lowest fluxes, the errors are overestimated by a factor of \(\sim\) 5--6. For
brighter objects, this difference declines until the re-estimated errors become
higher than those based on the RMS map.

\section{Rest-frame filter wavelength coverage}

In Table~\ref{tab:filters}, we show the restframe wavelength coverage of each UV
filter at the redshifts of each of the seven candidates. We are quoting the
wavelengths from 1\%--99\% of the integrated dimensionless throughput of each
filters as sampled in the software package \textsc{PySynPhot}
\citep{pysynphot2013}.

\begin{table}[tbhp]
  \caption{\label{tab:filters}Rest-frame UV filter wavelength coverage for each
    candidate.
  }
\centering
  \begin{tabular}{lrrrrrr}
  \toprule
  {} & \multicolumn{2}{l}{F225W} & \multicolumn{2}{l}{F275W} & \multicolumn{2}{l}{F336W} \\
  {} & $\lambda_{\text{min}}$ & $\lambda_{\text{max}}$ & $\lambda_{\text{min}}$ & $\lambda_{\text{max}}$ & $\lambda_{\text{min}}$ & $\lambda_{\text{max}}$ \\
  \midrule
  F225W-314  &     680 &     946 &     797 &    1019 &    1025 &    1218 \\
  F275W-2055 &     613 &     852 &     718 &     918 &     924 &    1097 \\
  F336W-189  &     458 &     636 &     536 &     685 &     690 &     819 \\
  F336W-554  &     527 &     733 &     618 &     790 &     795 &     944 \\
  F336W-606  &     547 &     761 &     641 &     820 &     825 &     980 \\
  F336W-1013 &     513 &     713 &     601 &     768 &     773 &     918 \\
  F336W-1041 &     471 &     655 &     552 &     706 &     710 &     844 \\
  \bottomrule
  \end{tabular}
\end{table}

\section{Artifact in PSF matched F336W image}
\label{sec:org0141b3f}

\begin{figure}[tbh]
\begin{center}
\includegraphics[width=0.99\columnwidth]{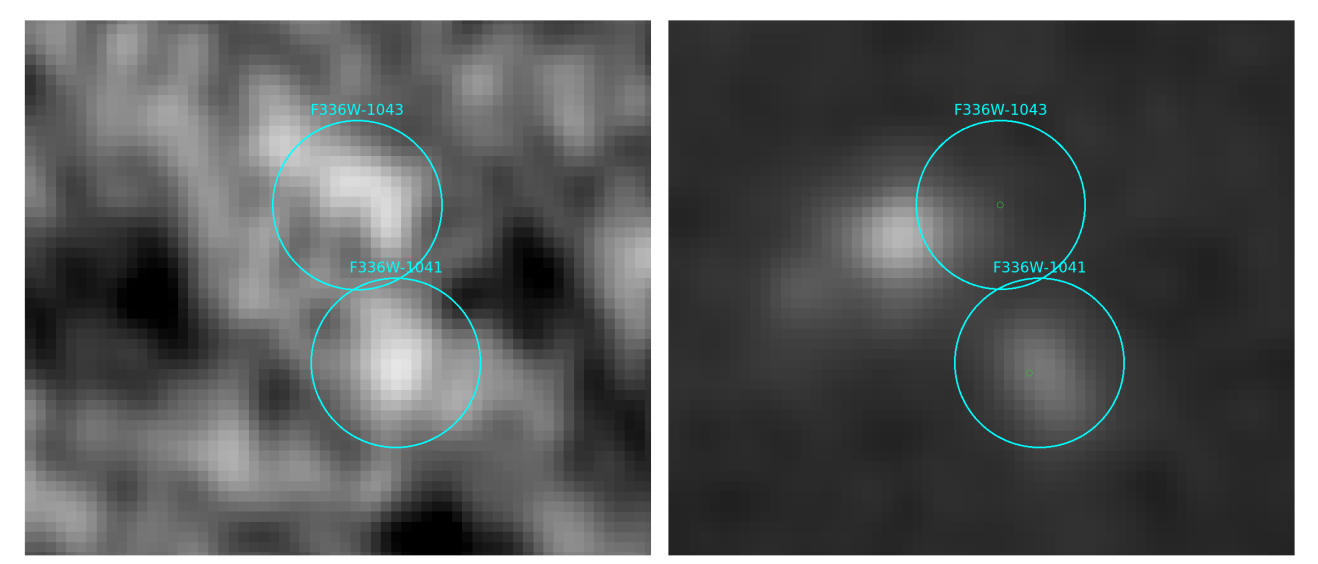}
\caption{\label{fig:mergerpair}Cutouts of F336W-1041 (lower circle) and F336W-1043 (upper circle) in the filters F336W (left panel) and F775W (right panel). The circles are of arbitrary size. The upper feature in F336W is visibly offset from the upper feature in F775W, which is true for the rest of optical and IR data. In contrast, the spatial coincidence of the lower feature in the two filters is clear.}
\end{center}
\end{figure}

Our source detection picked up an apparent feature in F336W \(\sim 1''\)
from F336W-1041. The feature is offset from the I17 and R15 catalog
coordinates of the companion galaxy and, as is visible in
fig.~\ref{fig:mergerpair}, also from the flux centroid in the optical
frames (here F775W). We inspected the non-PSF matched archival frame
as found in MAST and found that there was no source at that location
in that filter. We thus suspect that the feature is an artifact of the
PSF matching process.

To test whether the measured ionizing flux in 1041 could likewise be
an artifact, we extracted the flux from a circular aperture matching
the one used above and measured the flux and uncertainty. The values
we recovered for both flux and uncertainty were consistent with the
ones previously found, and its centroid in F336W coincident with that
in the optical filters, so we conclude that this feature was real.

\end{appendix} 

\end{document}